\documentclass[10pt,journal,compsoc]{IEEEtran}
\ifCLASSOPTIONcompsoc
  \usepackage[nocompress]{cite}
\else
  \usepackage{cite}
\fi
\ifCLASSINFOpdf
\else
\fi

\usepackage[numbers,sort&compress,square]{natbib}
\usepackage[autostyle]{csquotes}

\usepackage{cite}
\usepackage{url}
\usepackage{parskip}
\usepackage{indentfirst} %indents first line of each section
\usepackage{color}
\usepackage{graphicx}
\usepackage{caption}
\usepackage{subcaption}
\usepackage{diagbox}
\usepackage[linesnumbered,ruled,vlined]{algorithm2e}
\usepackage{flushend}
\usepackage{mathtools}
\usepackage{amsmath}
\usepackage{verbatim}
\usepackage{pbox}
\newenvironment{proof}[1][Proof]{\begin{trivlist}
\item[\hskip \labelsep {\bfseries #1}]}{\end{trivlist}}
\begin{document}
%\title{CAGE: A Market-based \underline{C}ontention-\underline{A}ware \underline{G}am\underline{E}-theoretic Distributed model for heterogeneous resource assignment}
\title{CARMA: Contention-aware Auction-based Resource Management in Architecture}

\author{
{\rm  Farshid Farhat, ~\IEEEmembership{Student Member,~IEEE}, Diman Zad Tootaghaj, ~\IEEEmembership{Student Member,~IEEE} }\\
\thanks{We thank, Novella Bartolini and Mohammad Arjomand for their feedback on earlier drafts of this paper. }
\thanks{D. Z. Tootaghaj and Farshid Farhat are with the Comp. Sci. Dept. in the Pennsylvania State University, PA, USA (email: {\{dxz149, fuf111\}@cse.psu.edu}). A partial and preliminary version appeared in Proc. IEEE ICCD'17 \cite{tootaghajICCD}.}
}


\markboth{}%IEEE Transactions on Emerging Topics in Computing}%, August~2015}%
{Shell \MakeLowercase{\textit{et al.}}: Bare Demo of IEEEtran.cls for Computer Society Journals}



\IEEEtitleabstractindextext{%
\begin{abstract}
\noindent As the number of resources on chip multiprocessors (CMPs) increases, the complexity of how to best allocate these resources increases drastically. Because the higher number of applications makes the interaction and impacts of various memory levels more complex. Also, the selection of the objective function to define what \enquote{best} means for all applications is challenging. Memory-level parallelism (MLP) aware replacement algorithms in CMPs try to maximize the overall system performance or equalize each application's performance degradation due to sharing. However, depending on the selected \enquote{performance} metric, these algorithms are not efficiently implemented, because these centralized approaches mostly need some further information regarding about applications' need. In this paper, we propose a contention-aware game-theoretic resource management approach (CARMA) using market auction mechanism to find an optimal strategy for each application in a resource competition game. The applications learn through repeated interactions to choose their action on choosing the shared resources. Specifically, we consider two cases: (i) cache competition game, and (ii) main processor and co-processor congestion game. We enforce costs for each resource and derive bidding strategy. Accurate evaluation of the proposed approach show that our distributed allocation is scalable and outperforms the static and traditional approaches.
%Traditional resource management systems rely on a centralized approach to manage users running on each resource. The centralized resource management system is not scalable for large-scale servers as the number of users running on shared resources is increasing dramatically and the centralized manager may not have enough information about applications' need. In this paper we propose a distributed game-theoretic resource management approach using market auction mechanism to find optimal strategy in a resource competition game. The applications learn through repeated interactions to choose their action on choosing the shared resources. Specifically, we look into two case studies of cache competition game and main processor and co-processor congestion game. We enforce costs for each resource and derive bidding strategy. Accurate evaluation of the proposed approach show that our distributed allocation is scalable and outperforms the static and traditional approaches.
\end{abstract}

\begin{IEEEkeywords}
Game Theory, Resource Allocation, Auction.
\end{IEEEkeywords}}

\maketitle

\IEEEdisplaynontitleabstractindextext

% \fi
\IEEEpeerreviewmaketitle

% or

\begin{figure}[!tb]
\centering
\includegraphics[height=1.5in, width=3.5in]{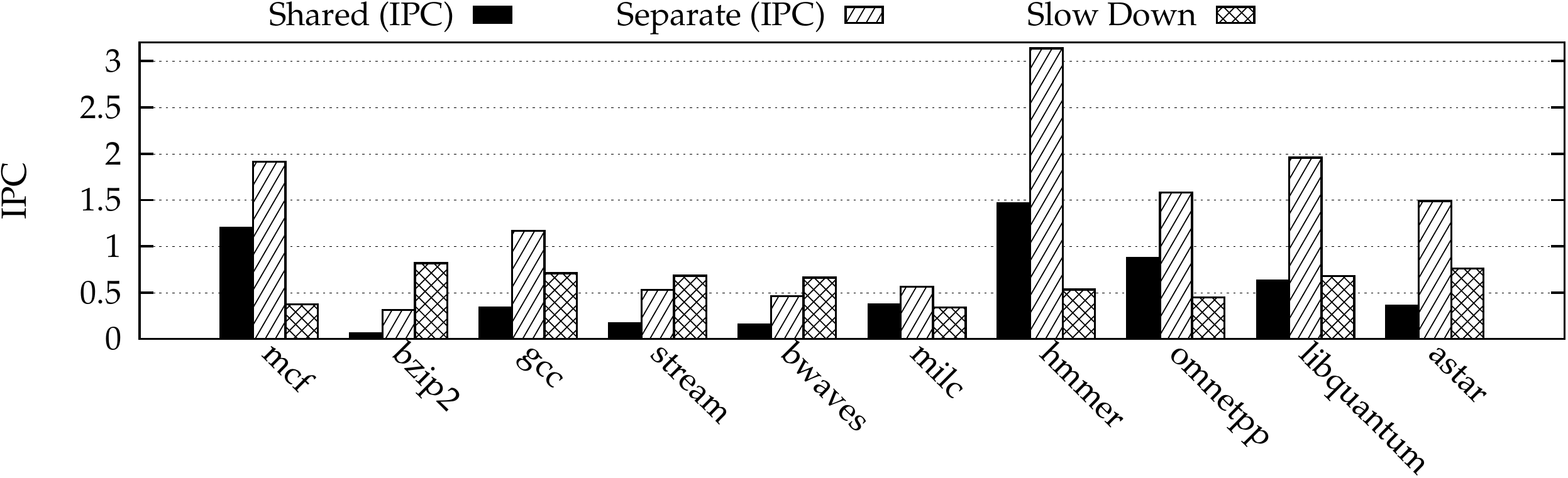} %Slow_down.pdf
\vspace{-1.5\baselineskip}
\caption{Performance degradation of 10 different \textit{spec 2006} applications sharing \textit{LLC}.}
\label{fig:Slow_down}
\vspace{-1.2\baselineskip}
\end{figure}
\begin{figure}[!b]
\vspace{-1.5\baselineskip}
\centering
\includegraphics[height=1.5in, width=3.5in]{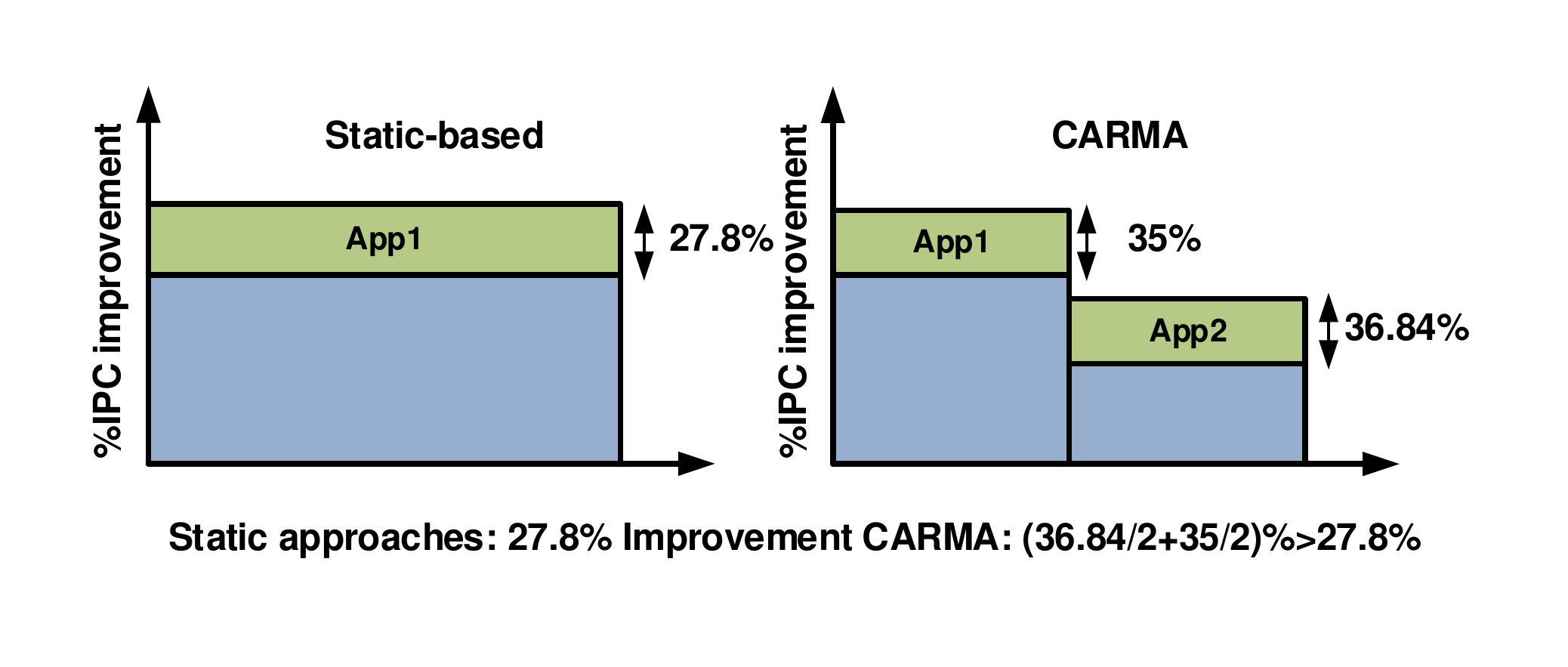} %Slow_down.pdf
\vspace{-2\baselineskip}
\caption{\label{fig:Dynamic} Performance comparison of static and dynamic scheduling of two applications (\textit{hmmer} and \textit{mcf} from \textit{Spec 2006}) under two different \textit{LLC} sizes.}
\end{figure}

%%%%%%%%%%%%%%%%%%%%%%%%%%%%%%%%%%%%%%%%%%%%%%%%%%%%%%%%%%%%%%%%%%%%%%%%%%%
%%%%%%%%%%%%%%%%%%%%%%%%%%%%%%%%%%%%%%%%%%%%%%%%%%%%%%%%%%%%%%%%%%%%%%%%%%%
\IEEEraisesectionheading{\section{Introduction} \label{introduction}}
\IEEEPARstart{T}{he} number of cores on chip multiprocessors (\textit{CMP}) is increasing each year and it is believed that only many-core architectures can handle the massive parallel applications. Server-side \textit{CMP}s usually have more than 16 cores and potentially more than hundreds of applications can run on each server. These systems are going to be the future generation of the multi-core processor servers. Applications running on these systems share the same resources like last level cache (\textit{LLC}), interconnection network, memory controllers, off-chip memories or co-processors where the higher number of applications makes the interaction and impacts of various resource levels more complex. Along with the rapid growth of core integration, the performance of applications highly depend on the allocation of the resources and especially the \textit{contention} for shared resources \cite{tang2011impact, zhuravlev2010addressing, hsu2006communist, kim2004fair, cho2006managing, tootaghajICCD, farhat2016stochastic, tootaghaj2016optimal, tootaghaj2015evaluating, farhat2016towardsStoc, tootaghaj2015thesis}. In particular, as the number of co-runners running on a shared resource increases, the magnitude of performance degradation increases. Also, the selection of the objective function to define what \enquote{best} means for all applications is challenging or even theoretically impossible to improve IPC of one application and memory latency of another application simultaneously in a system. As a result, this new architectural paradigm introduces several new challenges in terms of scalability of resource management and assignment on these large-scale servers. Therefore, a scalable competition method between applications to reach the optimal assignment can significantly improve the performance of co-runners on a shared resource. Figure~\ref{fig:Slow_down} shows an example of performance degradation for 10 \textit{spec 2006} applications when running on a shared 10MB \textit{LLC} (Shared), or when running on a private 1MB \textit{LLC} (Separate).  \\
%\indent Mohammad:\\
%%%%%%%%%%%%%%%%%%%%%%%%%%%%%%%%%%%%%%%%%%%%%%%%%%%%%%%%%%%%%%%%%%%%%%%%%%%%%
%%%%%%%%%%%%%%%%%%%%%%%%%%%%%%%%%%%%%%%%%%%%%%%%%%%%%%%%%%%%%%%%%%%%%%%%%%%%%
\indent Among these shared resources, sharing \textit{CPU}s and \textit{LLC}s plays an important role in overall CMP utilization and performance. Modern \textit{CMP}s are moving towards heterogeneous architecture designs where one can get advantage of both small number of high performance \textit{CPU}s or higher number of low performance cores. The advent \textit{Intel Xeon Phi} co-processors is an example of such heterogeneous architectures that during run-time the programmer can decide to run any part of the code on small number of \textit{Xeon} processors or higher number of \textit{Xeon Phi} co-processors. Therefore, the burden of making decisions on getting the shared resources is moving towards the applications. In addition to the shared \textit{CPU}s, shared \textit{LLC} keeps data on chip and reduces off-chip communication costs \cite{liu2004organizing}. Sometimes an application may flood on a cache and occupy a large portion of available memory and hurt performance of another application which rarely loads on memory, but its accesses are usually latency-sensitive. Recently, many proposals target partitioning the cache space between applications such that (1) each application gets the minimum required space, so that per-application performance is guaranteed to be at an acceptable level, (2) system performance is improved by deciding how the remaining space should be allocated to each one. \\
\indent Prior schemes \cite{zhuravlev2010addressing, qureshi2006utility, lin2008gaining, iyer2004cqos, liu2004organizing, rafique2006architectural, jiang2008analysis} are marching towards these two goals, usually by trading off the system complexity and maximum system utilization. It is shown that neither a pure private \textit{LLC}, nor a pure shared \textit{LLC}, can provide optimal performance for different workloads \cite{cho2006managing}. In general, cache partitioning techniques can be divided into way partitioning and co-scheduling techniques. In a set-associative cache, partitioning is done by per-way allocation. For example, in a 4-way 512KB shared cache allocating 128KB to application \textit{A} means to allow it storing data blocks in only one way per-set, without accessing remaining. Co-scheduling techniques try to co-schedule a set of applications with lowest interference together at the same time such that the magnitude of slow-down for each application is the same or a performance metric is optimized for all applications. However, it is shown that, depending on the objective function for the performance metric, cache allocation can result in totally different allocations \cite{hsu2006communist}. In general Prior schemes have the following three limitations:\\
%\begin{enumerate}
\indent \textbf{1. Scalability}: All of the prior schemes suffer from scalability; especially when the approach is tracking the application's dynamism \cite{qureshi2006utility, zhuravlev2010addressing, jiang2008analysis}. The reason is that algorithm complexity becomes higher in dynamic approaches. The root cause of this complexity is that all previous techniques make decisions (cache partitioning, co-scheduling) centralized using a central hardware or software. For example, main algorithm of \cite{qureshi2006utility} has exponential complexity $O( \binom{N+K-1}{K-1} )$ where $N$ is the number of applications sharing \textit{LLC} and $K$ is the number of ways. Table~\ref{table:complexity} shows the state of the art cache partitioning algorithms and their complexity of checking performance of different permutations.\\
\indent \textbf{2. Static-based}: Most of the prior works, use static co-scheduling to degrade slow-down of co-running applications on the same shared cache. However, static-based approaches cannot catch dynamic behavior of applications. Figure~\ref{fig:Dynamic} shows an example of two applications' IPC (\textit{hmmer} and \textit{mcf}) from \textit{Spec 2006} under different \textit{LLC} sizes. Let us consider a case where we have two cache sizes, a large cache of 1MB which can be shared between applications, and two private caches of 512KB which are not shared. The two applications are competing for the cache space. Suppose that both applications have two phases $(0,T)$ and $(T,2T)$. If the first application gets the larger cache space its \textit{IPC} increases by 35 percent in the first phase and by 20.6 percent in the second phase. The second application's \textit{IPC} increases by 15 percent in the first phase and by 36.84 percent in the second phase if it gets the larger cache space. In a static-based scheduling approach, the larger \textit{LLC} is always allocated to the first application with higher IPC in the time interval $(0, 2T)$, but in \textit{CARMA}, the applications compete for the shared resources, and in the first phase, the larger \textit{LLC} is allocated to the first application and in the second phase \textit{CARMA} allocates the larger \textit{LLC} to the second application. Therefore, static-based approaches cannot capture the dynamism in application's behavior and ultimately degrade the performance significantly. \\
\indent \textbf{3. Fairness}: Defining a single parameter for fairness is challenging for multiple applications, since applications have different performance benefits from each resource during each phase. In prior works fairness has been defined as a unique metric (eg. IPC, Power, Weighted Speed-up) for all applications. Therefore, in current approaches, the optimization goal of algorithms is the same for all applications. Consequently, we cannot sum up applications that desire different metrics in the same platform to decide on. However, if one application needs better IPC and another requires lower energy, the previous algorithms are not able to model it. The only way to address diversity of metrics (to be optimized) is to have an appropriate translation between different metrics (eg. IPC to Power) that is not trivial, while not addressed in prior study.\\
\begin{table}[!tb] %\scriptsize
\centering
\caption{\label{table:complexity} Complexity comparison of state-of-the-art \textit{LLC} partitioning/co-scheduling algorithms.}
\begin{tabular}{|c|c|} 
\hline Algorithm & Search Space\\
\hline Utility-based main algorithm \cite{qureshi2006utility} & $ \binom{N+K-1}{N-1}$  \\
\hline \pbox{20cm}{Greedy Co-scheduling \cite{jiang2008analysis}\\  $N$ applications and $N/K$ caches} & $ \binom{N}{K}  $ \\
\hline \pbox{20cm}{Hierarchical perfect matching \cite{jiang2008analysis} \\ $N$ applications} & $N^4 $ \\
\hline \pbox{20cm}{Local optimization \cite{jiang2008analysis} \\ $N$ applications and $N/K$ caches} & ${(N/K)}^2 \binom{2K}{K} $\\
\hline \pbox{20cm}{CARMA \\ $N$ applications and $K$ resources} & ${O(NK)}$\\
\hline
\end{tabular}
\vspace{-1\baselineskip}
\end{table}
\begin{comment}
\begin{figure}%[!htb]
\centering
%\includegraphics[height=3in, width=2.5in]{NodeArchs2.pdf}
\includegraphics[height=2in, width=3.5in]{Images/Complexity.pdf}
%\epsfig{file=Dataset.eps, height=2.5in, width=3in}
\caption{\label{fig:complexity} Complexity comparison of state-of-the-art \textit{LLC} partitioning algorithms.}
\end{figure}
\end{comment}
%%%%%%%%%%%%%%%%%%%%%%%%%%%%%%%%%%%%%%%%%%%%%%%%%%%%%%%%%%%%%%%%%%%%%%%%%%%%
%%%%%%%%%%%%%%%%%%%%%%%%%%%%%%%%%%%%%%%%%%%%%%%%%%%%%%%%%%%%%%%%%%%%%%%%%%%%
%\textcolor{red}{( Histogram charts on 128Kb cache for each and show how another 256Kb shared should be allocated to them on [Moin] and how it can be improved by dynamic decision (Figure 1 in STT-RAM paper).}\\
%\end{enumerate}
\indent In this paper, we present a game-theoretic resource assignment method to address all the above shortcomings including scalability, dynamism and fairness, while applications can get their desired performance based on their utility functions.\\
\indent \textbf{1. Semi-Decentralized}: Dual of each centralized problem is decentralized, if the optimization goal is broken into a smaller meaningful sub-problems. In the context of heterogeneous resource assignment this is straightforward. The profiling, analyzing and evaluating the demands are on application side, but the final decision on assigning the resources to applications based on the applications' bids is easily performed by the OS while they compete with each other for the best assignment. Like a capitalist system, the complexity of the governing transfers to the independent entities, and the government just make the policies and the final decisions. To achieve this, we introduce a novel market-based approach. Roughly speaking, the complexity of our approach in worst case scenario (for each application) is $O(NK)$ where $N$ is the number of the applications and $K$ is the number of available resources. However, on average the auction terminates in less than $N/2$ iterations.\\
\indent \textbf{2. Dynamic}: In order to confront the scalability problem of previous approaches, we use a market-based approach to move the decision making to the individual applications. Iterative auctions have been designed to solve non-trivial resource allocation problems with low complexity cost in government sale of resources, \textit{eBay}, real estate sales and stock market. Similarly, decentralized computation complexity is lower than centralized (for each application) which provides the opportunity to make the decision revisiting the allocation in small time quantum, or when a new application leaves or comes into the system.\\
\indent \textbf{3. Fair}: The proposed method solves the heterogeneous resource assignment problem in the context of marketing. Applications' demand regardless of the global optimization objective (IPC, Power, etc.) translates to the true valuation of their own performance. Resource assignment to the applications with the highest bids is performed by the auctioneer (the OS); making it local optimization objectives. Hence, resource assignment can be performed for different applications with different objectives known as utility functions.\\
\indent Overall, the proposed approach for cache contention game on average brings in 33.6\% improvement in system performance (when running 16 applications) compared to shared \textit{LLC}; while reaching less than 11.1\% of the maximum achievable performance in the best dynamic scheme. In the case study of heterogeneous CPU assignment, it brings in 106.6\% improvement (when running 16 applications at the same time). Also, the performance improvement increases even more as the number of co-running applications increases in the system. \\
\indent \textbf{Other potentials}: We introduce an auction-based resource management approach for different applications in large-scale competition games. In short, we as a system owner pay for a high-end CMP system for servers and guarantee that each application/user takes its best from the system by paying us back, or we as an application owner bid/pay the system to get the resources for my best performance. The auctioneer is application-agnostic, and does not interfere with applications' profile to globally optimize the system, but the applications compete for their own improvement. The two case studies of cache partitioning and CPU sharing are examples for resource sharing and the proposed approach can be employed in other resource partitioning algorithms. \\
\indent The reminder of the paper is organized as follows. Section~\ref{Motivation} discusses the background and motivation behind this work. In section~\ref{Problem_definition}, we discuss our auction-based game model. Section~\ref{Case_Studies} discusses the case study of cache contention game and the case study of main processor and co-processor contention and simulation results. Section~\ref{Related_works} studies related works and Section~\ref{Conclusion} concludes the paper with a summary.

\vspace{-0.5\baselineskip}
\section{Motivation and Background} \label{Motivation}
\subsection{Motivation}
\begin{figure}[!tb]
\centering
\includegraphics[height=1.8in, width=3.5in]{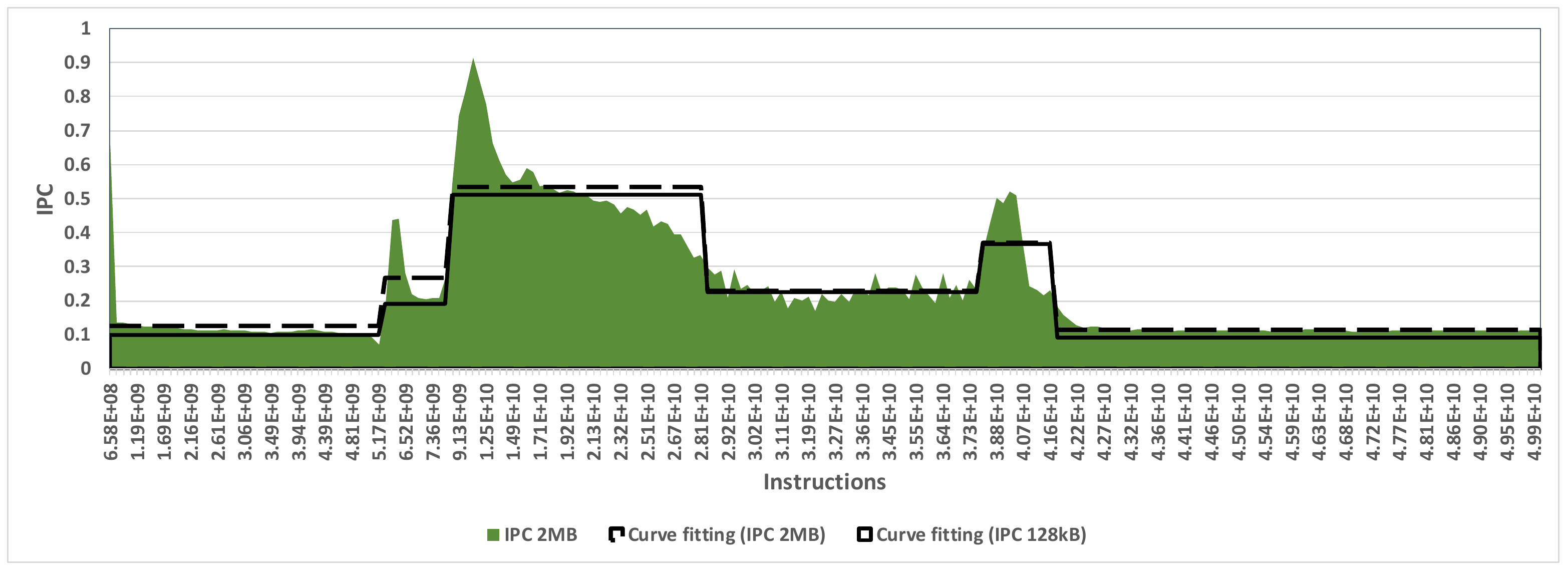} %Phases.pdf
\vspace{-1.5\baselineskip}
\caption{\label{fig:Phases}Phase transition in mcf with different L2 cache sizes.}
\vspace{-1.0\baselineskip}
\end{figure}

Different applications have different resource constraint with respect to CPU, memory, and bandwidth usage. Having a single resource manager for all existing resources and users in the system result in inefficiencies since it is not scalable and the operating system may not have enough information about applications' needs. For example, traditional LRU-based cache strategy uses cache utilization as a metric to give larger cache size to the applications which have higher utilization and lower cache size to the applications with lower cache utilization. However more cache utilization does not always result in better performance. Streaming applications for example have very high cache utilization, but very small cache reuse. In fact, the streaming applications only need a small cache space to buffer the streaming data. With rapid improvements in semiconductor technology, more and more cores are being embedded into a single core and managing large scale application using a single resource manager becomes more challenging. \\
%\indent Even if the applications are forced to announce their resource demand, it is possible that they lie about their resource vector or run some useless instructions to pretend to utilize the allocated resources given to them.
\indent In addition, defining a single fairness parameter for multiple applications is non-trivial since applications have different bottlenecks and may get different performance benefits from each resources during each phases of their execution time. Defining a single reasonable parameter for fairness is somewhat problematic. For instance, simple assignment algorithms which try to equally distribute the resources between all applications ignores the fact that different applications have different resource constraints. As a consequence, this makes the centralized resource management systems very inefficient in terms of fairness as well as performance needs of applications. We need a decentralized framework, where all applications' performance benefit could be translated into a unique notion of fairness and performance objective (known as utility function in economics) and the algorithm tries to allocate resources based on this translated notion of fairness. This translation has been well defined in economics and marketing, where the diversity of customer needs, makes more economically efficient market \cite{zhou2014sharing}. Economists have shown that in an economically efficient market, having diverse resource constraints and letting the customers compete for the resources can make a Nash equilibrium where both the applications and the resource managers can be enriched. Furthermore, applications' demand changes over time. Most resource allocation schemes pre-allocate the resources without considering the dynamism in applications' need and number of users sharing the same resource over time. Therefore, applications' performance can degrade drastically over time. Figure~\ref{fig:Phases} shows phase transitions for instruction per cycle (IPC) of mcf application from \textit{spec 2006} over 50 billion instructions. \\
%%%%%%%%%%%%%%%%%%%%%%%%%%%%%%%%%%%%%%%%%%%%%%%%%%%%%%%%%%%%%%%%%%%%%%%%%%%
%%%%%%%%%%%%%%%%%%%%%%%%%%%%%%%%%%%%%%%%%%%%%%%%%%%%%%%%%%%%%%%%%%%%%%%%%%%
\indent We try to find a game-theoretic distributed resource management approach where the shared hardware resources are exposed to the applications and we show that by running a repeated auction game between different applications which are assumed to be rational, the output of the game converges to a balanced Nash equilibrium allocation. In addition, we compare the convergence time of the proposed algorithm in terms of dynamism in the system. We evaluate our model with two case studies: 1) Private and shared last level cache problem, where the applications have to decide if they would benefit from a larger cache space which can potentially get more congested or a smaller cache space which is potentially less congested. 2) Heterogeneous processors (\textit{Intel Xeon} and \textit{Xeon Phi}) problem, where we perform experiments to show how congestion affects the performance of different applications running on an \textit{Intel Xeon} or \textit{Xeon Phi} co-processors. Depending on the amount of congestion in the system, the application can offload the most time consuming part of its code on the \textit{Xeon Phi} co-processors or not.
%%%%%%%%%%%%%%%%%%%%%%%%%%%%%%%%%%%%%%%%%%%%%%%%%%%%%%%%%%%%%%%%%%%%%%%%%%%
\vspace{-1\baselineskip}
\subsection{Background}
%Congestion games have been studied in network routing protocols where the delay of each player choosing a path in the network depends on the number of players choosing the same route in the system. 
%Every congestion game is a potential game since there exists a potential function associated with it. In addition, every congestion game has a pure-strategy Nash equilibrium. A key assumption in congestion games is that all users have the same impact on the congestion. However, this assumption is not always true. In case of computer architecture resources, applications effect each other differently and dividing the payoff function by the number of users running on the shared resource does not give us the correct utility. 
Game theory has been used extensively in economics, political and social decision making situations \cite{tootaghaj2011game, tootaghaj2011risk, kotobi2017spectrum, kotobi2015introduction, kesidis2013distributed, kurve2013agent, wang2017using, wang2015recouping}. A game is a situation, where the output of each player not only depends on her own action in the game, but also on the action of other players \cite{osborne1994course}. Auction games are a class of games which has been used to formulate real-world problems of assigning different resources between $n$ users. Auction game framework can model resource competition, where the payoff (cost) of each application in the system is a function of the contention level (number of applications) in the game.\\
\indent Inspired by market-based interactions in real life games, there exists a repeated interaction between competitors in a resource sharing game. Assuming large number of applications, we show that the system gets to a Nash equilibrium where all applications are happy with their resource assignment and don't want to change their state. Furthermore, we show that the auction model is \textit{strategy-proof}, such that no application can get more utilization by bidding more or less than the true value of the resource. In this paper we propose a distributed market based approach to enforce cost on each resource in the system and remove the complexity of resource assignment from the central decision maker.\\ 
\indent The traditional resource assignment is performed by the operating system or a central hardware to assign fair amount of resources to different applications. However, fair scheduling is not always optimal and solving the optimization problem of assigning $m$ resources between $n$ users in the system is an integer programming which is an NP-hard problem and finding the best assignment problem becomes computationally infeasible. Prior works focus on designing a fair scheduling function that maximizes all application's benefit \cite{zahedi2014ref, llull2017cooper, ghodsi2011dominant, zahedi2015sharing, fan2016computational}, while applications might have completely different demands and it is not possible to use the same fairness function for all. By shifting decision making to the individual applications, the system becomes scalable and the burden of establishing fairness is removed from the centralized decision maker, since individual applications have to compete for the resources they need. Applications start by profiling the utility function for each resource and bid for the most profitable resource. During the course of execution time they can update their belief based on the observed performance metrics at each round of the auction. Updating the utility functions at each round of the auction is based on the history of the observed performance metrics which shows the state of the game. 
%The idea behind updating the utility functions is that the history at each round of the auction shows the state of the game. 
This state indicates the contention on the current acquired resources. The payoff function in each round depends on the state of the system and on the action of other applications in the system. 
%%%%%%%%%%%%%%%%%%%%%%%%%%%%%%%%%%%%%%%%%%%%%%%%%%%%%%%%%%%%%%%%%%%%%%%%%%%
\subsubsection{Sequential Auction}
Auction-based algorithms are used for maximum weighted perfect matching in a bipartite graph $G=(U,V, E)$ \cite{bertsekas1998network, kyle1985continuous, vasconcelos2009bipartite}. A vertex  $U_i \in U$ is the application in the auction and a vertex $V_j \in V$ is interpreted as a resource. The weight of each edge from $U_i$ to $V_j$ shows the utility of getting that particular resource by $U_i$. The prices are initially set to zero and will be updated during each iteration of the auction. In sequential auctions, each resource is taken out by the auctioneer and is sequentially auctioned to the applications, until all the resources are sold out.
\subsubsection{Parallel Auction}
In a parallel auction, the applications submit their bids for the first most profitable item. The value of the bid at each iteration is computed based on the difference of the highest profitable object and the second highest profitable object. The auctioneer would assign the resources based on the current bids. At each iteration, the valuation of each resource is updated based on the observed information during run-time which shows the contention on that particular resource.
%%%%%%%%%%%%%%%%%%%%%%%%%%%%%%%%%%%%%%%%%%%%%%%%%%%%%%%%%%%%%%%%%%%%%%%%%%%%
%%%%%%%%%%%%%%%%%%%%%%%%%%%%%%%%%%%%%%%%%%%%%%%%%%%%%%%%%%%%%%%%%%%%%%%%%%%%
%\vspace{-1\baselineskip}
\section{The Method}\label{Problem_definition}
\begin{figure*}[!htb]
\centering
\includegraphics[height=3.2in, width=6.5in]{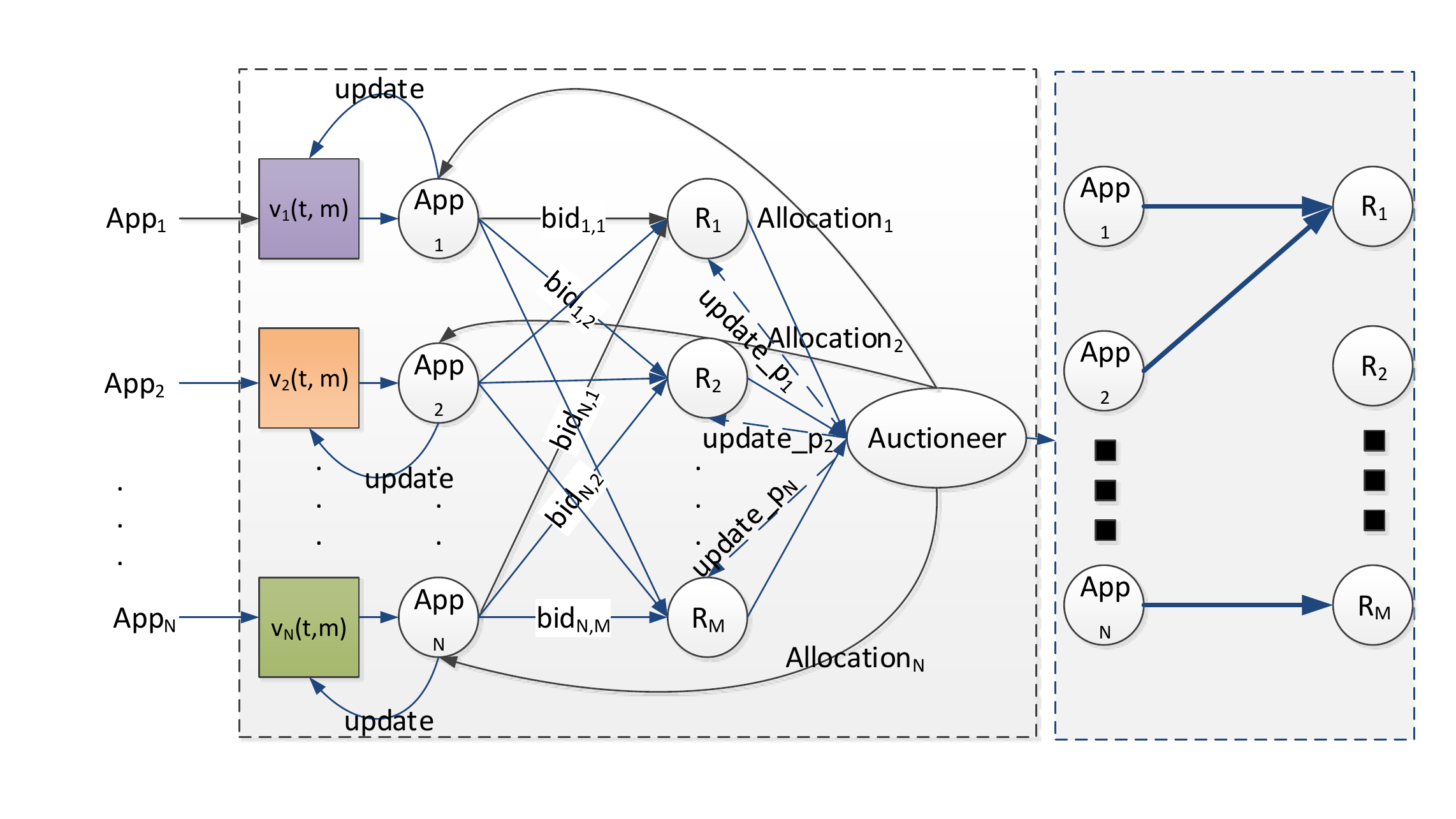} %[height=4in, width=8in]
\vspace{-1\baselineskip}
\caption{\label{fig:auction} Framework for auction-based resource assignment (CARMA).}
\vspace{-0.5\baselineskip}
\end{figure*}

Consider $n$ applications and $i$ instances of $m$ different resources. Applications arrive in the system one at a time. The applications have to choose among $m$ resources. There exists a bipartite graph between the matching of the applications and the resources.\\
\indent In general, there can be more than one application to get a shared resource. However, each application cannot get more than one of the available heterogeneous resources. For example, if we have two cache spaces of size 128kB (one way) and 256kB (two ways), each application can either get the 128kB, or the 256kB cache space and can't get both of them at the same time. Furthermore, each resource $R_k$ has a cost $p_k$ which is defined by the applications' bid in the auction. \\
\indent Figure~\ref{fig:auction} shows our auction-based framework to support \textit{CARMA} between $N$ applications that execute together competing for $M$ different resources. Each application has a utility table that shows how much performance it gets from each $M$ resources at each time slot. Based on the utility tables, applications submit bids for the most profitable resource. Based on the submitted bids, the auctioneer decides about the resource assignment for each resource, and updates the prices. Next, the applications who did not get any assignment compete for the next most profitable resource based on the updated prices repeatedly until all applications are assigned.  Figure~\ref{fig:auction} shows an example of a resource assignment and the corresponding bipartite graph.
%Table~\ref{table:notation} shows the notation used in our formulation.
%%%%%%%%%%%%%%%%%%%%%%%%%%%%%%%%%%%%%%%%%%%%%%%%%%%%%%%%%%%%%%%%%%%%%%%%%%%%%%%
%%%%%%%%%%%%%%%%%%%%%%%%%%%%%%%%%%%%%%%%%%%%%%%%%%%%%%%%%%%%%%%%%%%%%%%%%%
\begin{table}[!tb] \scriptsize
\centering
\caption{Notation used in our formulations.}\label{Table:notation}
\begin{tabular}{|p{0.7in}|p{2.3in}|} 
\hline $N$ & Number of players or applications (from $App_1$ to $App_N$). \\
\hline $M$ & Number of resources (from $R_1$ to $R_M$). \\
\hline $\vec{m}$ & A positive $M\times 1$ vector in the resource space that shows how much each application gets from each resource. \\   % of size $M\time 1$ showing the amount for each resources. \\ Number of applications which can get a resource \\ 
% \hline $n$ & Number of applications competing for a specified resource \\
\hline $T$ & Time intervals where the bidding is held \\
\hline $t_{i,j}$ & $j$-th phase time for $i$-th application during its course of execution time. \\ 
\hline $T_i$ & Last phase time for application $i$. \\
\hline $v_{i}(t,\vec{m})$ & The valuation function of application $i$ for the resource assignment $\vec{m}$ at time $t$. \\
\hline $v_{i,j}(t,\vec{m},r)$ & The valuation function of (application $i$,resource $j$), if we replace the $j$-th resource in the resource vector $m$ by $r$ \\
\hline $\delta$ & dynamic factor that shows how much we can rely on the past iterations. \\
\hline $G=(U,V,E)$ & A bipartite graph showing the resource allocation between the applications and the set of resources. \\
\hline $U$ & The set of applications which shows the left set of nodes in the bipartite graph $G=(U,V,E)$. \\
\hline $V$ & The set of resources which shows the right set of nodes in the bipartite graph $G=(U,V,E)$. \\
\hline $E$ & The edges in the bipartite graph. \\
\hline $b_{i,k}$ & User i's bid for k-th resource. \\
\hline $F_i$ & The total budget (summation of bids) a user have. \\
\hline $C_k$ & The total capacity of each resource. \\
\hline $p_{k}$ & The price of resource $k \in V$ in the auction. \\
%\hline $Bottleneck_{1,i}$ & The first bottleneck resource for application $i$ \\
%\hline $Bottleneck_{2,i}$ & The second bottleneck resource for application $i$ \\
\hline $K$ & Number of cache levels \\
\hline
\end{tabular}
\vspace{-1.0\baselineskip}
\end{table}
%%%%%%%%%%%%%%%%%%%%%%%%%%%%%%%%%%%%%%%%%%%%%%%%%%%%%%%%%%%%%%%%%%%%%%%%%%%%%%%
%%%%%%%%%%%%%%%%%%%%%%%%%%%%%%%%%%%%%%%%%%%%%%%%%%%%%%%%%%%%%%%%%%%%%%%%%%%
\begin{comment}
\begin{figure}[!htb]
\centering
%\includegraphics[height=3in, width=1.5in]{NodeArchs2.pdf}
\includegraphics[height=2.2in, width=1.3in]{Images/bipartite.pdf}
%\epsfig{file=Dataset.eps, height=2.5in, width=3in}
\caption{\label{fig:bipartite} Cache allocation as a bipartite graph.}
\end{figure}
\end{comment}
%%%%%%%%%%%%%%%%%%%%%%%%%%%%%%%%%%%%%%%%%%%%%%%%%%%%%%%%%%%%%%%%%%%%%%%%%%%
\vspace{-1\baselineskip}
\subsection{Problem Defenition} 
\indent We formulate our problem as an auction to enforce cost/value updates for each resource as follows: 
%The cost of each player to get a resource is the cost of the assigned resource divided by the number of players who share. 
\begin{itemize}
  \item \textbf{Valuation $\mathbf{v_{i}(t,\vec{m})}$}: Application $i$ has a valuation function which shows how much it benefits from the resource vector $\vec{m}$ at time $t$. The valuation function at time $t=0$ for cache contention case study is derived from the IPC (instruction per cycle) curves using profiling, and for processor and co-processor contention case study is derived from the profiling of separate cache performance of the application. However, in general, each application can choose its own utility function.
  %%%%%%%%%%%%%%%%%%%%%%%%%%%%%%%%%%%%%%%%%%%%%%%%%%%%%% 
    \item \textbf{Observed information}: The observed information at each time step is the performance value of the selected action in the game. Therefore, the applications repeatedly update the history of their valuation function over time.  
  %%%%%%%%%%%%%%%%%%%%%%%%%%%%%%%%%%%%%%%%%%%%%%%%%%%%%%   
    \item \textbf{Belief updating}: Let $T$ be the time intervals where the bidding is held. At each iteration step of the auction, the applications update their valuation of each resource based on the observed performance on the resource vector. The update at time $W$ is derived using the following formula:
%\begin{small}
\begin{equation}\label{eq:belief}
v_{i}(W,\vec{m})=\frac{\sum\limits_{0\leq n\leq W/T} {\delta}^{W/T-n} \cdot v_{i}(nT,\vec{m})}{\sum\limits_{0\leq n\leq W/T} {\delta}^{W/T-n}},
\end{equation}
%\end{small}  
%%%%%%%%%%%%%%%%%%%%%%%%%%%%%%%%%%%%%%%%%%%%%%%%%%%%%%%%%%%%%%%%%%%%%%%%%
where $v_{i}(W,\vec{m})$ shows the observed valuation of resource vector $\vec{m}$ at time $W$ by application $i$ in the system; $\delta$ shows the discount factor between 0 and 1 which shows how much a user relies on its past observations in the system. The discount factor is chosen to show the dynamics of the system. If the observed information in the system changes fast, the discount factor is close to zero, i.e. the application cannot rely on the past observations very much. However, if the system is more stable and the observed information does not change fast, the discount factor is closer to 1. If a user fails in an auction, its payoff and corresponding observed valuation at the current time is equal to zero. So, it won't probably bid for the same resource vector again, since its valuation decreases for next round. We choose the discount factor to be the absolute value of the correlation coefficient of the observed values of the valuations at each iteration step which is calculated as follows:
%\begin{small}
\begin{equation}
\delta =  \frac{E(v_{i}(W,\vec{m}))^2}{{\sigma_{v_{i}(W,\vec{m})}}^2}
\end{equation}  
%\end{small}
%%%%%%%%%%%%%%%%%%%%%%%%%%%%%%%%%%%%%%%%%%%%%%%%%%%%%%%%%%%%%%%%%%%%%%%%
%%%%%%%%%%%%%%%%%%%%%%%%%%%%%%%%%%%%%%%%%%%%%%%%%%%%%%%%%%%%%%%%%%%%%%%%
  \item \textbf{Action}: At each time step, the applications decide which resource to bid and how much to bid for each resource. 
\end{itemize} 
%%%%%%%%%%%%%%%%%%%%%%%%%%%%%%%%%%%%%%%%%%%%%%%%%%%%%%%%%%%%%%%%%%%%%%%%%% 
\indent Table~\ref{Table:notation} shows important notation used throughout the paper. In the following sections, we describe our distributed optimization scheme to solve the problem. 
%%%%%%%%%%%%%%%%%%%%%%%%%%%%%%%%%%%%%%%%%%%%%%%%%%%%%%%%%%%%%%%%%%%%%%%
%\begin{equation}
%min \sum\limits_{i=1}^n v_i C_k \frac{b_{i,k}}{\theta_k}, \\
%s.t. \sum\limits_{i=1}^n b_{i,k} \leq E_i
%\end{equation}
%%%%%%%%%%%%%%%%%%%%%%%%%%%%%%%%%%%%%%%%%%%%%%%%%%%%%%%%%%%%%%%%%%%%%%%
\vspace{-1\baselineskip}
\subsection{Distributed Optimization Scheme}
The goal is to design a repeated auction mechanism which runs by the operating system to guide the applications to choose their best resource allocation strategy. The applications' goal is to maximize their own performance and the operating system wants to maximize the total utility gain from the applications. Each application can use its own utility function and evaluates the resources based on the desired value of the resources. \\
\indent \textbf{Applications' approach}: The application $i$ wants to maximize the expected utility (pay-off) with respect to a limited budget ($F_i$) during all phases of its execution time. We have:
%\vspace{-0.5\baselineskip}
%%%%%%%%%%%%%%%%%%%%%%%%%%%%%%%%%%%%%%%%%%%%%%%%%%%%%%%%%%%%%%%%%%%%%%%%%
%maximize \;\;\;\; \sum\limits_{i=1}^n v_i C_k \frac{b_{i,k}}{\theta_k},\\
%\begin{small}
\begin{align}
%\begin{IEEEeqnarray}{rCl}
\forall i \in U \; \; \; \; \; maximize \; \; \; \; \sum\limits_{0<t<T_i} v_{i}(t,\vec{m})-b_{i}(t,\vec{m}) , \nonumber \\
 % \IEEEyessubnumber\\
subject \; to \;\;\;\; \sum\limits_{0<t<T_i} b_{i}(t,\vec{m}) \leq F_i, \; \; \; \; \forall i \in U.
%\IEEEyessubnumber
%\end{IEEEeqnarray}
\end{align}
%\end{small}
%%%%%%%%%%%%%%%%%%%%%%%%%%%%%%%%%%%%%%%%%%%%%%%%%%%%%%%%%%%%%%%%%%%%%%%%%%
\indent \textbf{OS's approach}: The operating system wants to maximize the social welfare function which is translated into submitted bids from the applications in a limited resource constraints.
\vspace{-0.5\baselineskip}
%%%%%%%%%%%%%%%%%%%%%%%%%%%%%%%%%%%%%%%%%%%%%%%%%%%%%%%%%%%%%%%%%%%%%%%%%%
%\begin{small}
\begin{align}
%\begin{IEEEeqnarray}{rCl}
maximize \; \; \; \sum\limits_{i=1}^N \sum\limits_{0<t<T_i}  b_{i}(t,\vec{m}) \cdot A_{i}(t, \vec{m}) , \nonumber \\ 
%\IEEEyessubnumber\\
subject \; to \;\;\;\; \sum\limits_{i=1}^N  A_{i}(t, \vec{m}) \leq A_{max}, \; \; \; \; \forall t: 0 \leq t \leq T , \nonumber \\
%\IEEEyessubnumber\\
A_{i}(t, \vec{m}) \in \{0,1\} , \; \; \forall i \in U, \; \;  \forall \vec{m} \subseteq \mathcal{V}, \; \;  \forall t:  0 \leq t \leq T ,
\end{align}
where the binary variable $A_{i}(t, \vec{m})$ represents the decision to assign resource vector $\vec{m}$ to application $i$ at time $t$ (when $A_{i}(t, \vec{m}) =1$) or not (when $A_{i}(t, \vec{m})=0$); and $\mathcal{V}$ is the vector space of the all resource vectors ($\forall\vec{m}$); and $A_{max}$ shows the maximum number of the applications which can share the resource vector $\vec{m}$. 

%%%%%%%%%%%%%%%%%%%%%%%%%%%%%%%%%%%%%%%%%%%%%%%%%%%%%%%%%%%%%%%%%%%%%%%%%%%
%\begin{comment}
\indent \textbf{Illustrative example}: As an illustrative example shown in Figure~\ref{fig:Dynamic}, let us consider a case where we have two different resources, a large cache of 1MB which can be shared between applications, and two private caches of 512KB which are not shared. 
%There are two applications competing for the cache space. One of the applications wants to minimize its request latency and the other one wants to maximize number of instructions executed per cycle. Suppose that both applications have two phases $(0,T)$ and $(T,2T)$. If the first application gets the larger cache space its \textit{IPC} increases by 15 percent in first phase and by 36.84 percent in the second phase. The second application's \textit{IPC} increases by 35 percent in the first phase and by 20.6 percent in the second phase if it gets the larger cache space. %Also, suppose that they both have 60 tokens as a budget to submit. 
The first application participates in the auctions with 35¢ bid at the first phase and 21¢ bid at the next phase. The second application participates in the auctions with 15¢ bid at the first phase and 37¢ bid at the next phase. The auctioneer (OS) decides to allocate larger cache in the auctions at first phase to the first application and at next phase to the second application. %Since, the social welfare would be maximized if the auctioneer allocates both applications with the larger cache space, they would both get the larger resource. Then the first application notices that its utility function does not improve as he predicts and adjusts the utility table and can either change its allocation or stay on current allocation. 
%The application should redistribute the tokens for the next phase if it did not get the desired resource in the first phase. The second application invests 35 tokens in the first phase and 25 tokens in the next phase. The auctioneer (OS) at each phase decides to allocate which resource to which applications. Since, the social welfare would be maximized if the auctioneer allocates both applications with the larger cache space, they would both get the larger resource. Then the first application notices that its utility function does not improve as he predicts and adjusts the utility table and can either change its allocation or stay on current allocation. 
%If both applications bid 10\$ for the private cache and 15\$ for the shared cache, the operating system would allocate both the shared cache space and get 15\$ from each to maximize its revenue.
%\end{comment}
\vspace{-1\baselineskip}
\subsection{Analysis}
The distributed optimization problem is hard to solve. However, in reality, the problem can split into simpler subproblems since each application knows its bottleneck resource and would first bid for the first bottleneck resource to maximize its utility.\\
\indent We suppose all applications in the system are risk-neutral which means they have a linear valuation of the utility function. Each risk-neutral agent wants to maximize its expected revenue. Risk attitude behaviors are defined in \cite{ferber1999multi} where the agents can broadly be divided into risk-averse, risk-seeking and risk neutral. Risk-averse agents prefer deterministic values rather than risky value profits and risk-seeking applications have a super-linear utility function and prefer risky utilities than sure utilities. Next, we derive the Bayes Nash equilibrium strategy profile for all agents in the system assuming risk neutrality. 
%%%%%%%%%%%%%%%%%%%%%%%%%%%%%%%%%%%%%%%%%%%%%%%%%%%%%%%%%%%%%%%%%%
\newtheorem{defi}{Definition}
\begin{defi}
A strategy profile $a$ is a pure Nash equilibrium if for every application $i$ and every strategy $a_i' \neq a_i \in A$ we have $u_i(a_i, a_{-i}) \geq u_i(a_i', a_{-i})$
\end{defi}
\newtheorem{theorem}{Theorem}
\begin{theorem}\label{thm:neat}
%\emph{(Theorem)}
\label{Auction}
Suppose $n$ risk-neutral applications whose valuations are derived uniformly and independently from the interval $[0,1]$ compete for one resource which can be assigned to $m$ applications who have the highest bid in the auction. We will show that Bayes Nash equilibrium bidding strategy for each application in the system is to bid $\frac{n-m}{n-m+1}v_i$ where $v_i$ is the profit of application $i$ for getting the specified resource.  
\end{theorem}
%%%%%%%%%%%%%%%%%%%%%%%%%%%%%%%%%%%%%%%%%%%%%%%%%%%%%%%%%%%%%%%
%\vspace{-1\baselineskip}
\begin{algorithm}[!tb] %\small
\DontPrintSemicolon % Some LaTeX compilers require you to use \dontprintsemicolon    instead
\KwIn{A bipartite Graph (U, V, E).}
\KwOut{The allocation of resources to applications.}
The initial resource vector for each application is the average amount across various resources. At time $t=nT$, the valuation of each application for each resource vector is updated using Eq.~\ref{eq:belief}.

For application $U_i \in U$, the first bottleneck resource is
%\[ Bottleneck_{1,i} = V_{i,m}=  arg \; \max_{m \in V} (v_{i,m}-p_{m})  \] 
%\[ Bottleneck_1: V_{i,j^{max}_1}=  \max_{1\leq j_1 \leq M} {\Delta v_{i,j_1}(t,\vec{m},r_{j_1})-p_{j_1}}  \] 
\[ V_{i,j^{max}_1}=  \max_{1\leq j_1 \leq M} {\Delta v_{i,j_1}(t,\vec{m},r_{j_1})-p_{j_1}}  \] 
where the differential valuation function is $\Delta v_{i,j}(t,\vec{m},r_j) = v_{i,j}(t,\vec{m},r_j) - v_{i}(t,\vec{m})$. 

Find the second bottleneck resource for application $U_i \in U$ in the system:
%\[ Bottleneck_2: V_{i,j^{max}_2}=  \max_{1\leq j_2 \leq M; j_2\neq j_1} {\Delta v_{i,j_2}(t,\vec{m},r_{j_2})-p_{j_2}}  \] 
\[ V_{i,j^{max}_2}=  \max_{1\leq j_2 \leq M; j_2\neq j_1} {\Delta v_{i,j_2}(t,\vec{m},r_{j_2})-p_{j_2}}  \] 

Each application calculates the partial bid for its first bottleneck resource using the following formula:
\[ b_{i,j^{max}_1}(t) = V_{i,j^{max}_1} - V_{i,j^{max}_2} + p_{j^{max}_1} + \epsilon \]

Each resource $r_j \in V$ which can be shared between $L$ applications, is assigned to the $L$ highest bidding applications $Winner_{i,j}=\{i_1, i_2, ..., i_L\}$ and the price for that resource is updated as follows:
%\[ p_{j} = \max_{i_1, i_2, ..., i_L \in U} \sum\limits_{k=1}^L {b_{i_k,j}}  \]
\[ p_{j} = \max_{l \in \{1,...,L\}} {b_{i_l,j}}  \]

The $minBid$ for each resource is updated as the minimum bid of $l$ applications who acquired $j$-th resource. That is:
\[ B^{min}_j=  \min_{i_l \in Winner_{i,j}} {b_{i_l,j}} \] 

Goto step 2 until all partial bids for all resources are determined. 

The total bid of the application $i$ is as follows:
\[ b_{i}(t,\vec{m})=b_{i}(t,[r_{j_1};r_{j_2};...;r_{j_M}])=\sum\limits_{j=1}^M {b_{i,j}(t)} \]
where iteratively $\vec{m}=[r_{j_1};r_{j_2};...;r_{j_M}]$.

Find the estimated investment $I_{i}(t)$ using Algorithm~\ref{algo:p} to plan the upper bound of the investment with respect to the budget $F_i$. If $I_{i}(t)\geq b_{i}(t,\vec{m})$, application $i$ will participate in this auction at time $t$, otherwise it quits and other applications do the steps 2 and 3.

\caption{Parallel Auction for Heterogeneous Resource Assignment}
\label{algo:b}
\end{algorithm}
\begin{algorithm}[!tb] %\small
\DontPrintSemicolon % Some LaTeX compilers require you to use \dontprintsemicolon    instead
\KwIn{A bipartite Graph (U, V, E).}
\KwOut{Participation (YES) in an auction or Quit (NO).}
At time $t=nT$, assume that we have the same state in terms of resources.

For application $U_i \in U$, we similarly run the steps 2, 3, 4, 5 and 6 of Algorithm~\ref{algo:b} to find all estimated bids in next rounds based on its various phases. We have:
\[ b_{i}(t_{i,j},\vec{m})=\sum\limits_{j=1}^M {b_{i,j}(t)};\forall t_{i,j}>t \]
Also, we have the previous bids of the application $i$: 
\[ b_{i}(t_{i,j},\vec{m});\forall t_{i,j}<t \]

If $F_i \geq \sum\limits_{\forall t_{i,j}\neq t} {b_{i}(t_{i,j},\vec{m})}$, then YES and the application will participate in the auction. Otherwise NO, and the application will update the zero valuation for current round using Eq.~\ref{eq:belief}.

\caption{Budget Planning}
\label{algo:p}
\vspace{0\baselineskip}
\end{algorithm}
%%%%%%%%%%%%%%%%%%%%%%%%%%%%%%%%%%%%%%%%%%%%%%%%%%%%%%%%%%%%%%%%%%%%%%%%%%%%%
\indent Theorem~\ref{thm:neat}, states that whenever there is a single resource that users compete to get it with different valuation functions, the Nash equilibrium strategy profile for risk-neutral users is to bid $\frac{n-m}{n-m+1}v_i$. This term tends to the true value of the object when n is a large number. \\
\indent In case of more than one resource competition we derive Algorithm~\ref{algo:b} for heterogeneous resource assignment and will prove that it has a Nash equilibrium in the game. Algorithm~\ref{algo:b} uses Algorithm~\ref{algo:p} to do budget planning for our purpose. In the first step, all valuations are set to the solo-run of application's performance. Next, each application submits a partial bid for its first bottleneck resource. The partial bid should be larger than the price of the object which is initialized to zero at the beginning of the program. The applications only have the incentive to bid a value that is no more than the difference between the first and second bottleneck resource. Otherwise, it submits a smaller bid to the second bottleneck and gets the same revenue as paying more for the first bottleneck resource. In order to break the equal valuation function between two different applications, we use $\epsilon$ scaling such that at each iteration of the auction the prices should increase by a small number. The OS will set the resources' price with these partial bids, and find the minimum of the highest partial bids for each resource. The applications recurse for all the resources, and the total bid is the summation of the partial bids for each application. Then, the applications execute Algorithm~\ref{algo:p} to participate in the auction or not. Finally the participated applications with the bids higher than $B^{min}_j$ will get $j$-th resource.\\
%In addition, suppose we have 5 different memory bandwidth exposed to the applications. Each application gets a different performance benefit from different cache sizes and different memory bandwidth which is denoted in table  **. The applications need to submit their bids based on their performance benefits. 
%\begin{equation}
%\begin{split}
%\int_0^\frac{nb_1}{n-m}  .... \int_0^\frac{nb_1}{n-m} \! (\frac{v_1}{m} -b_1) \, \mathrm{d}u_2 %\mathrm{d}u_3 ... \mathrm{d}u_{n-m}= \\
%= {(\frac{nb_1}{n-m}) }^{n-m} (\frac{v_1}{m} -b_1). 
%\end{split}
%\end{equation}
%%%%%%%%%%%%%%%%%%%%%%%%%%%%%%%%%%%%%%%%%%%%%%%%%%%%%%%%%%%%%%%%%%%%%%%%%%%%
%\[ \frac{\partial}{\partial b_1} ({(\frac{nb_1}{n-m}) }^{n-m} (\frac{v_1}{m} -b_1))
%\newtheorem{defi}{Definition}
%\begin{defi}
%Let's assume each user $Ui$ in the system is defined as one vertice of a graph $G$ in the system and let each edge in the graph shows which subset of users can impact each others' performance and the associated weight of each edge show the cost function of how two users affect each others' performance in the system. Each edge has a weight function denoted by ${P1(n), P2(n), ... Pe(n)}$, where $e$ is the number of edges in Graph $G$. Let $A=A_1 \times A_2 \times ... \times A_n $ be the set of actions that each user can play. 
%\end{defi}
\indent The overhead of the auction for the auctioneer (the OS) is very negligible. The OS during the auction only sets the prices of the resources based on the received bids from the applications and gives the resources to the highest bids. So every $T$ seconds, the OS runs these two jobs, which adds a negligible overhead with respect to other tasks of the OS. Our approach also satisfies the following properties:
1) Individual rationality (IR): Applications' expected utility is non-negative because the amount of the bid cannot be beyond the sum of the difference of the valuations which is at most the highest valuation of the application.
2) Truthfulness: Applications cannot benefit from bidding other than their true valuation. By contradiction, if an application bids lower than the true value, there may be another application with a higher bid to take the resource. But we cannot guarantee the truthfulness in the case of collusion among applications.
3) Budget-balance: The whole payments from the applications are less than the OS revenue, which is trivial as we have only one seller which is the OS.
4) Economic efficiency: It has been shown in \cite{bertsekas1998network} that this assignment is optimal, but it doesn't mean it is economically efficient since we know that it depends on the applications' valuation which is sub-optimal.

%\vspace{-1\baselineskip}
\section{Case Studies} \label{Case_Studies}
%\subsection{A case study of the main processor (Xeon) and coprocessor (Xeon-Phi) congestion game}
\subsection{CPU Scale-up Scale-out Game}
\begin{figure}[!tb]
\centering
\includegraphics[height=1.5in, width=3.5in]{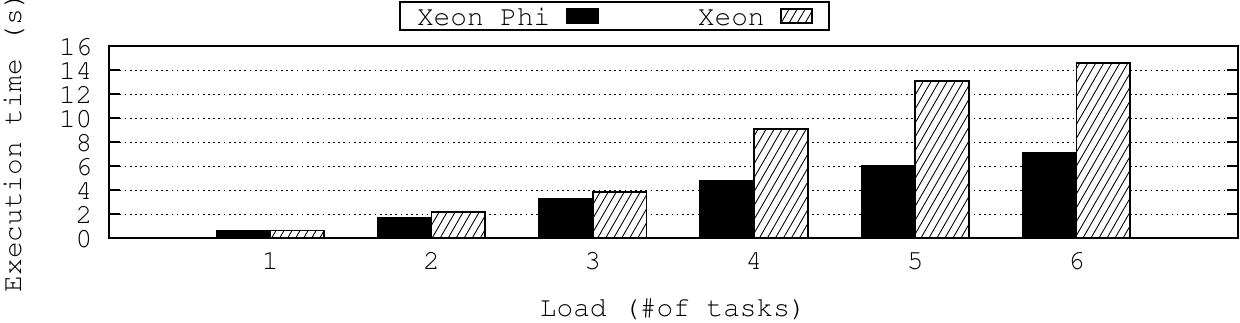} % diman.pdf
\caption{Congestion effect on \textit{Xeon} and \textit{Xeon Phi} machines.}
\label{Fig:congestion}
\vspace{-1\baselineskip}
\end{figure}

\begin{figure}[!tb]
\centering
\includegraphics[height=1.5in, width=3.5in]{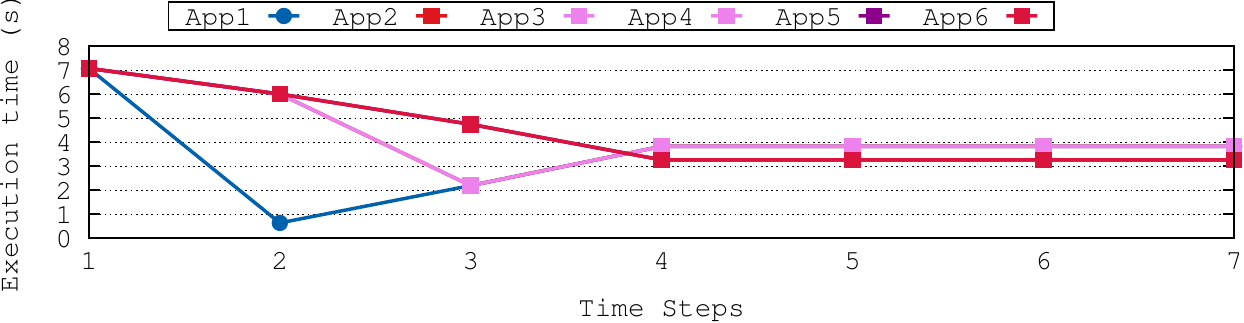}%Game_during_time.pdf
\caption{Performance of 6 instances of applications during the execution time for our proposed game model.}\label{Fig:performance_over_time}
\end{figure}

\begin{figure}[!tb]
\centering
\includegraphics[height=1.5in, width=3.5in]{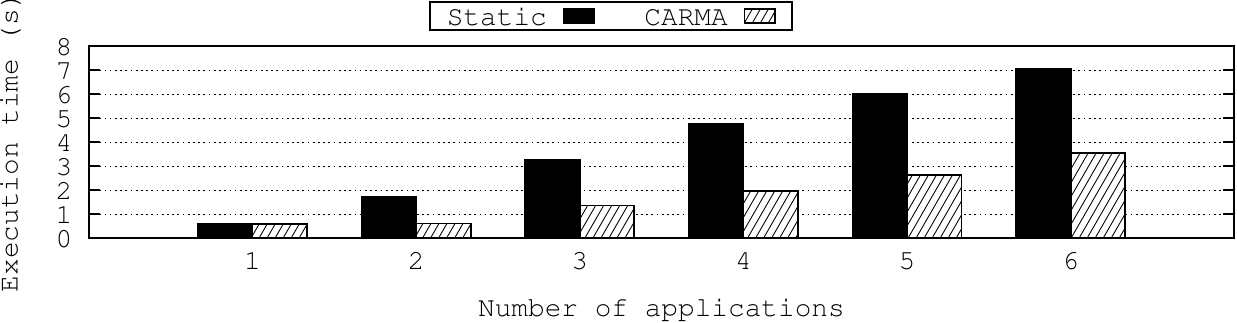}  %Congestion_aware.pdf
\caption{Performance comparison of congestion-aware schedule versus static schedule.}
\label{Fig:Perfomance_Comparison}
\vspace{-1\baselineskip}
\end{figure}

The emerging high-performance computing applications lead to the advent of \textit{Intel Xeon Phi} co-processor, that when their highly parallel architecture is fully utilized, can run in order of magnitude more performance than the existing processor architectures. The \textit{Xeon Phi} co-processors are the first commercial product of Intel \textit{MIC} processors where the hardware architecture is exposed to the programmer to choose running the code on either \textit{Xeon} processor or \textit{Xeon Phi} co-processors. It is possible that, during the course of execution, either the processor or the co-processor get congested and the performance of the application degrades a lot. Therefore, making a decision to offload the most time-consuming part of the program on \textit{Xeon} or \textit{Xeon Phi} should be made online, based on the contention level.  In this section, we look at the case study of our auction-based model on decision making of running the application on the main or co-processor in a highly congested environment. \\
\indent The experimental results of this section are run on \textit{Stampede} cluster of \textit{Texas Advanced Computing Center}. Table~\ref{Table:Xeon} shows the comparison of \textit{Intel Xeon} and \textit{Xeon Phi} architectures which is used in this section. 
%%%%%%%%%%%%%%%%%%%%%%%%%%%%%%%%%%%%%%%%%%%%%%%%%%%%%%%%%%%%%%%%%%%%%%%%%%%
%%%%%%%%%%%%%%%%%%%%%%%%%%%%%%%%%%%%%%%%%%%%%%%%%%%%%%%%%%%%%%%%%%%%%%%%%%%
%%%%%%%%%%%%%%%%%%%%%%%%%%%%%%%%%%%%%%%%%%%%%%%%%%%%%%%%%%%%%%%%%%%%%%%%%%%
\begin{table}[!tb] 
\centering
\caption{The comparison of \textit{Intel Xeon} processor and \textit{Intel Xeon Phi} processor.}\label{Table:Xeon}
\begin{tabular}{|c|p{0.8in}|p{1in}|} 
\hline Processors & Xeon E5-2680 & Xeon Phi SE10P \\
\hline Cores/Sockets & 8/2 & 61/1 \\
\hline Clock Frequency & 2.7 GHz & 1.1 GHz  \\
\hline Memory & 32GB 8x4G 4-channels DDR3-1600MHz & 8GB GDDR5 \\
\hline L1 cache & 32 KB & 32 KB \\
\hline L2 cache & 256 KB & 512 KB \\
\hline L3 cache & 20 MB & - \\
\hline
\end{tabular}
\end{table}
%%%%%%%%%%%%%%%%%%%%%%%%%%%%%%%%%%%%%%%%%%%%%%%%%%%%%%%%%%%%%%%%%%%%%%%%%%%
%%%%%%%%%%%%%%%%%%%%%%%%%%%%%%%%%%%%%%%%%%%%%%%%%%%%%%%%%%%%%%%%%%%%%%%%%%%
%%%%%%%%%%%%%%%%%%%%%%%%%%%%%%%%%%%%%%%%%%%%%%%%%%%%%%%%%%%%%%%%%%%%%%%%%%%
It is observed that congestion has a significant impact on the performance of running the application on \textit{Xeon} and \textit{Xeon Phi} machines. Since most cloud computing machines are shared between thousands of users, the programmer not only should get the benefit of parallelism by offloading the most time-consuming part of the code to the larger number of low-performance cores (\textit{Xeon Phi}) but also should consider the congestion level (number of co-runners) in the system. To this end, we performed experiments on \textit{Stampede} clusters. We executed \textit{MiniGhost} application which is a part of \textit{Mantevo} project \cite{mantevo} which uses difference stencils to solve partial differential equations using numerical methods. The applications use the profiling utility functions at $t=0$ and during the course of execution update the utility function based on the observed performance on each core using Equation~\ref{eq:belief}. Then, they can revisit their previous action on running the code on either the processor or co-processor during run-time. \\
\indent Figure~\ref{Fig:congestion} shows the total execution time with respect to congestion we made in \textit{Xeon} and \textit{Xeon Phi}. In this experiment we ran the same problem size on a \textit{Xeon} and \textit{Xeon Phi} machine multiple times so that we could see the effect of load on the total execution time of our application. It was observed that with the same number of threads \textit{Xeon}'s performance degrades more than \textit{Xeon phi}. 
Next, we tried to change the application behavior using a congestion-aware game theoretic algorithm to offload the most time-consuming part of the application based on the performance behavior of applications. Figure~\ref{Fig:performance_over_time} shows the result of our game-theoretic model during the execution time. It is observed that during the course of execution, the applications change their strategy on either choosing the main processor or the co-processor and all applications' performance converge to an equilibrium point where applications don't want to change their strategy. \\
\indent Furthermore, it is shown that CARMA can bring in up to 106.6\% improvement in total execution time of applications compared to static approach when the number of co-runners is six. The performance improvement would be significant when the number of co-runners increase. Figure~\ref{Fig:Perfomance_Comparison} shows the performance comparison of CARMA and static approach which does not consider the congestion dynamism in the system and the decision is only made based on the parallelism level in the code. 
%%%%%%%%%%%%%%%%%%%%%%%%%%%%%%%%%%%%%%%%%%%%%%%%%%%%%%%%%%%%%%%%%%%%%%%%%%%
%%%%%%%%%%%%%%%%%%%%%%%%%%%%%%%%%%%%%%%%%%%%%%%%%%%%%%%%%%%%%%%%%%%%%%%%%%%
%%%%%%%%%%%%%%%%%%%%%%%%%%%%%%%%%%%%%%%%%%%%%%%%%%%%%%%%%%%%%%%%%%%%%%%%%%%
%%%%%%%%%%%%%%%%%%%%%%%%%%%%%%%%%%%%%%%%%%%%%%%%%%%%%%%%%%%%%%%%%%%%%%%%%%%
%%%%%%%%%%%%%%%%%%%%%%%%%%%%%%%%%%%%%%%%%%%%%%%%%%%%%%%%%%%%%%%%%%%%%%%%%%%
%%%%%%%%%%%%%%%%%%%%%%%%%%%%%%%%%%%%%%%%%%%%%%%%%%%%%%%%%%%%%%%%%%%%%%%%%%%
%%%%%%%%%%%%%%%%%%%%%%%%%%%%%%%%%%%%%%%%%%%%%%%%%%%%%%%%%%%%%%%%%%%%%%%%%%%
%%%%%%%%%%%%%%%%%%%%%%%%%%%%%%%%%%%%%%%%%%%%%%%%%%%%%%%%%%%%%%%%%%%%%%%%%%%
%%%%%%%%%%%%%%%%%%%%%%%%%%%%%%%%%%%%%%%%%%%%%%%%%%%%%%%%%%%%%%%%%%%%%%%%%%%
%%%%%%%%%%%%%%%%%%%%%%%%%%%%%%%%%%%%%%%%%%%%%%%%%%%%%%%%%%%%%%%%%%%%%%%%%%%
\subsection{A Case Study of Private and shared cache game}
\begin{figure}[!tb]
\centering
\includegraphics[height=1.5in, width=3.3in]{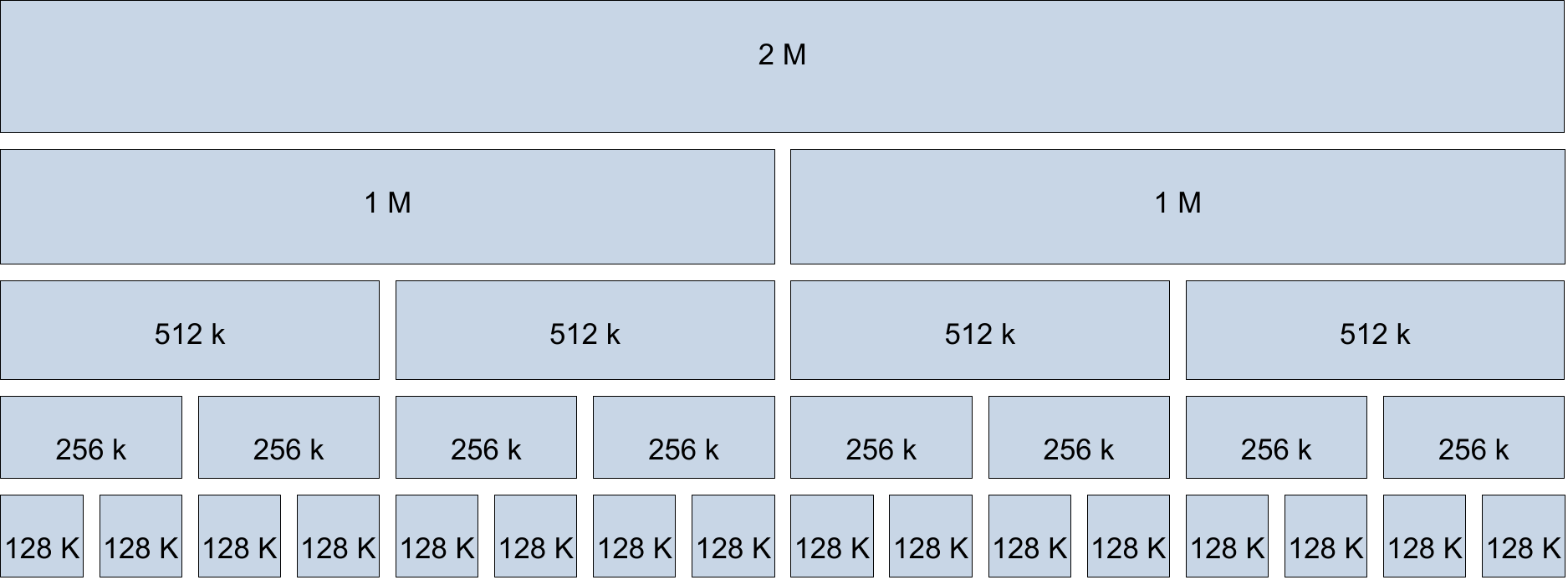}
\caption{The proposed last level cache hierarchy model.}\label{Fig:cache_hierarchy}
\end{figure}

\begin{figure*}[!tb]
\centering
\includegraphics[height=3.5in, width=6.5in]{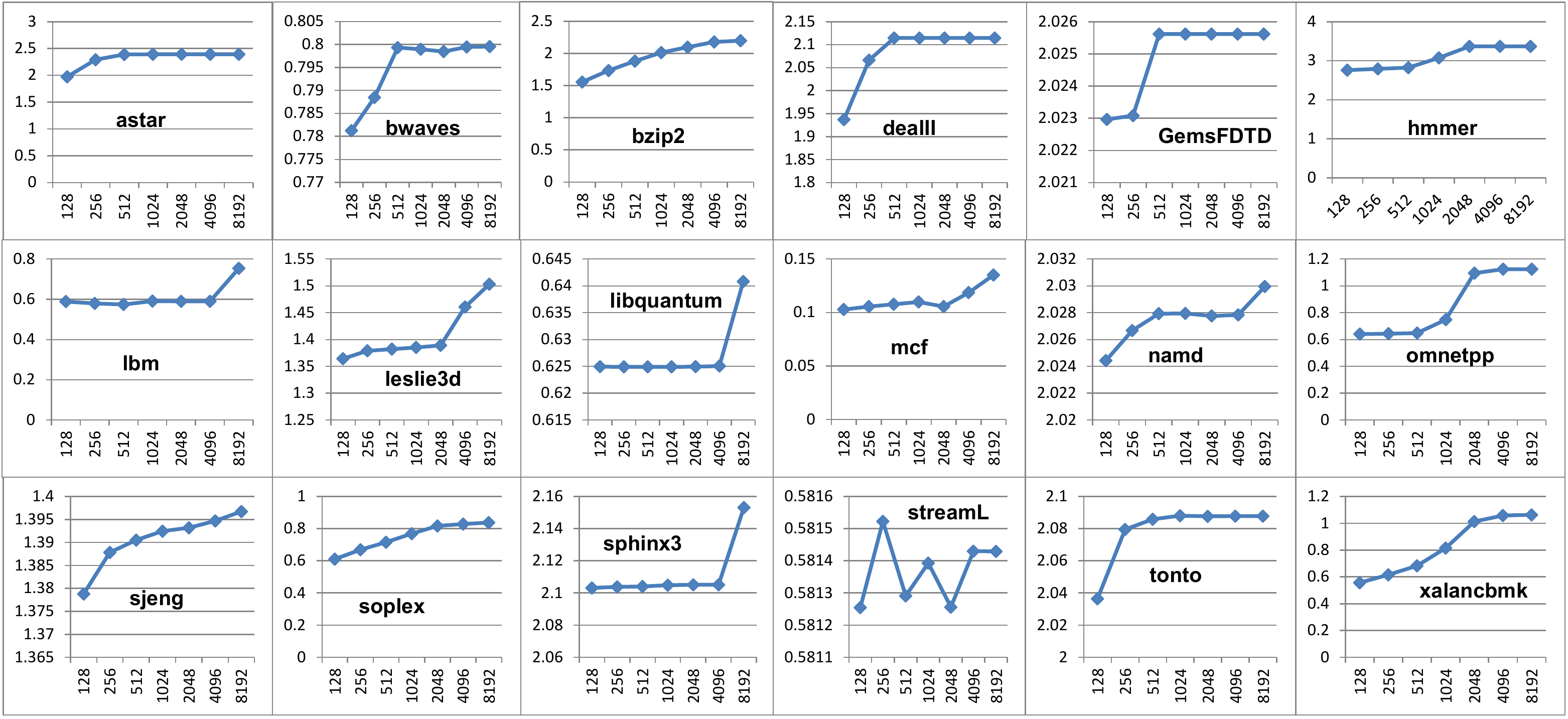}
\caption{IPC for different size of LLC.}\label{fig:Cache_IPC}  
\end{figure*}

\begin{figure*}[!htb]
        \centering
        \begin{subfigure}[b]{0.28\textwidth} %//0.28 bood
                \includegraphics[width=\textwidth]{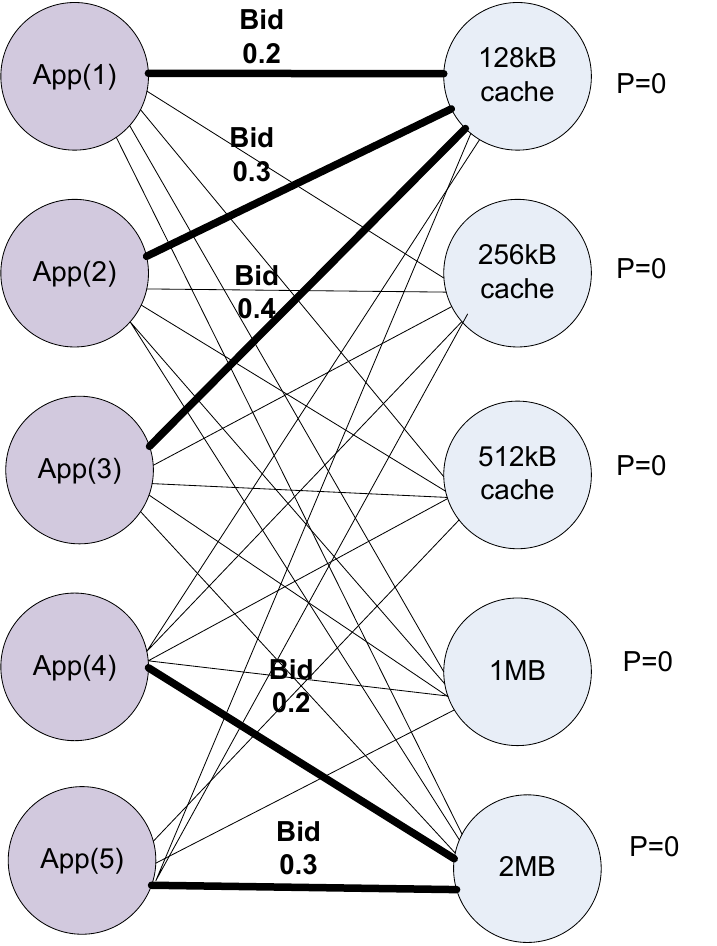}
                \caption{first round.}
                \label{fig:first_round}
        \end{subfigure}%
        ~ %add desired spacing between images, e. g. ~, \quad, \qquad etc.
        \begin{subfigure}[b]{0.28\textwidth}
                \includegraphics[width=\textwidth]{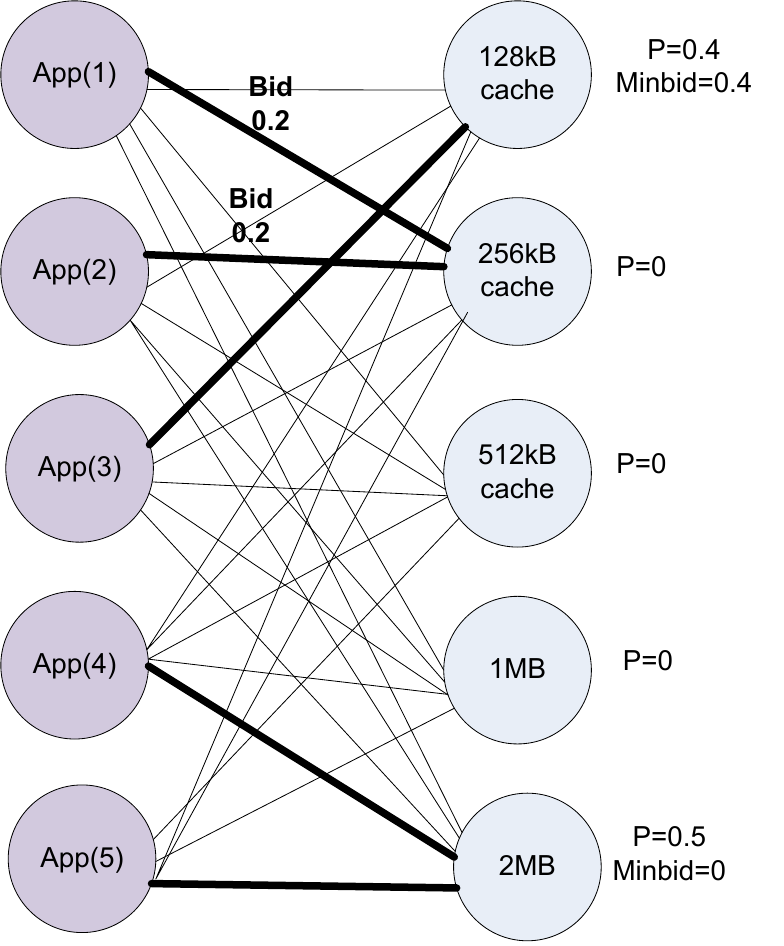}
                \caption{second round.}
                \label{fig:second_round}
        \end{subfigure}
        ~ %add desired spacing between images, e. g. ~, \quad, \qquad etc.
        \begin{subfigure}[b]{0.28\textwidth}
                \includegraphics[width=\textwidth]{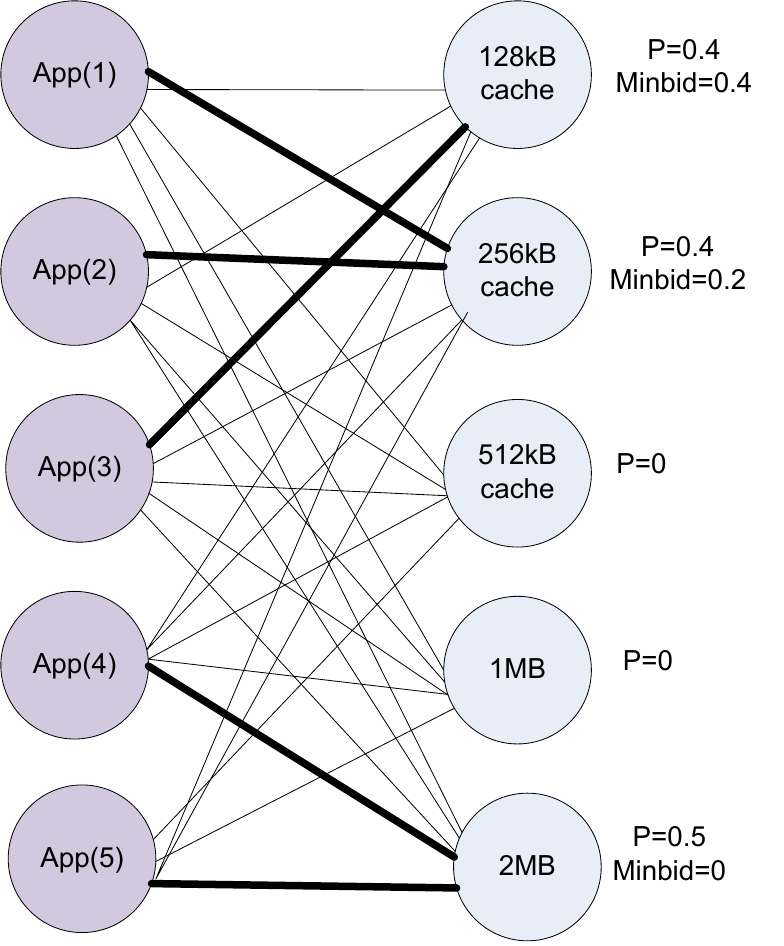}
                \caption{third round.}
                \label{fig:third_round}
        \end{subfigure}  

                \caption{Cache allocation, a) first round, b) second round and c) third round of bidding.}\label{fig:Auction_rounds}    
       % \vspace{-2\baselineskip}
\end{figure*}

One of the challenging problems in \textit{CMP} resource management systems is whether applications benefit from a shared large last level cache or an isolated private cache. We evaluated CARMA's performance, on a 10MB LLC shown in Figure~\ref{Fig:cache_hierarchy}, where 2MB, 1MB, 512kB, 256kB and 128kB levels of LLC can potentially be shared between 16, 8, 4, 2 and 1 applications respectively, the cache levels have 16, 8, 4, 2, and 1 ways. Table~\ref{Table:Workloads} summarizes the studies workloads and their characteristics, including miss per kilo instructions (\textit{MPKI}), memory bandwidth usage, and IPC. We use applications from \textit{Spec 2006} benchmark suite \cite{Spec:website}. We use \textit{Gem5} full system simulator in our experiment ~\cite{binkert2011gem5, Gem5:website}. Table~\ref{Table:Experimental_Set_Up} shows the experimental setup in our experiments.\\
\indent To evaluate the performance of our proposed approach we use utility functions for different number of cache ways shown in Figure~\ref{fig:Cache_IPC}. These utility functions at the start of the execution can be found using either profiling techniques or stack distance profile \cite{kim2004fair, suh2002new, suh2004dynamic} of applications assuming there are no co-runners in the system. Next, during run-time, the applications can update their utility functions based on Equation~\ref{eq:belief}. Therefore, there is a learning phase where applications learn about the state of the system and update the utilities accordingly. The stack distance profile indicates how many more cache misses will be added if the application has less number of ways in the cache. Based on the stack distance profile, the applications can update their utility function and bid for the next iteration of the auction if they like to change their allocation. Next, we bring an example of the auction for one time step of the game. This time step can be repeated once an application arrives or leaves the system or when an application's phase changes during run-time. However, in case of one application's phase change or arriving or leaving the system, the algorithm reaches the optimal assignment in much fewer iterations since all other assignments are fixed and a few applications would be affected.\\
%%%%%%%%%%%%%%%%%%%%%%%%%%%%%%%%%%%%%%%%%%%%%%%%%%%%%%%%%%%%%%%%%%%%%%%%%%%
%%%%%%%%%%%%%%%%%%%%%%%%%%%%%%%%%%%%%%%%%%%%%%%%%%%%%%%%%%%%%%%%%%%%%%%%%%%
%%%%%%%%%%%%%%%%%%%%%%%%%%%%%%%%%%%%%%%%%%%%%%%%%%%%%%%%%%%%%%%%%%%%%%%%%
\begin{table}[!tb] %\scriptsize
\centering
\caption{Experimental Setup.}
\label{Table:Experimental_Set_Up}
\begin{tabular}{|c|p{1.5in}|} 
\hline Processors & Single threaded with private L1 instruction and data caches \\
\hline Frequency & 1GHz \\
\hline L1 Private ICache & 32 kB, 64-byte lines, 4-way associative\\
\hline L1 Private DCache & 32 kB, 64-byte lines, 4-way associative \\
\hline L2 Shared Cache & 128 kb-2 MB, 64-byte lines, 16-way associative \\
\hline RAM & 12 GB \\
\hline
\end{tabular}
\end{table}
%%%%%%%%%%%%%%%%%%%%%%%%%%%%%%%%%%%%%%%%%%%%%%%%%%%%%%%%%%%%%%%%%%%%%%%%%
\begin{table}[!tb] 
\centering
\caption{Evaluated workloads.}
\label{Table:Workloads}
\begin{tabular}{p{0.7cm} p{1.5cm} p{1cm} p{1.7cm} p{1cm} }
\hline
%\begin{tabular}{|l|c|c|c|} \hline
{\bf \#} & \bf Benchmark & MPKI & Memory BW & IPC  \\
\hline 
{\bf 1} & astar & 1.319 & 373 MB/s & 2.057 \\
{\bf 2} & bwaves & 10.47 & 1715 MB/s & 0.661 \\
{\bf 3} & bzip2 & 3.557 & 1194 MB/s & 1.367 \\ 
{\bf 4} & dealII &  0.935 & 307 MB/s & 2.107 \\
{\bf 5} & GemsFDTD & 0.004 & 2.19 MB/s & 2.023 \\
{\bf 6} & hmmer & 2.113 & 1547 MB/s & 2.861 \\
{\bf 7} & lbm & 19.287 & 3954 MB/s & 0.533 \\
{\bf 8} & leslie3d & 8.469 & 1942 MB/s & 1.297 \\
{\bf 9} & libquantum & 10.388 & 1589 MB/s & 0.531 \\
{\bf 10} & mcf & 16.93 & 820 MB/s & 0.073 \\
{\bf 11} & namd & 0.051 & 20.32 MB/s & 2.362\\
{\bf 12} & omnetpp & 10.34 & 1147 MB/s & 0.504 \\
{\bf 13} & sjeng & 0.375 & 139.2 MB/s & 1.403 \\
{\bf 14} & soplex & 4.672 & 390.8 MB/s & 0.513 \\
{\bf 15} & sphinx3 & 0.349 & 202.8 MB/s & 2.223 \\
{\bf 16} & streamL & 31.682 & 3619 MB/s & 0.581 \\
{\bf 17} & tonto & 0.260 & 107 MB/s & 2.036 \\
{\bf 18} & xalancbmk & 12.703 & 1200 MB/s & 0.558 \\
%\hline  
\hline
\end{tabular}
\end{table}
%%%%%%%%%%%%%%%%%%%%%%%%%%%%%%%%%%%%%%%%%%%%%%%%%%%%%%%%%%%%%%%%%%%%%%%%%%
\begin{comment}
Assume $n$ different applications denoted by ${U1, U2, .. Un}$ with different cache benefits which may affect each other with different cost functions. Figure 1 shows the different applications which are the vertices of the graph with their impact on each other which are the weights of edges in the graph. If two vertices are not connected in the graph, it means that they would not affect each others' performance. For example, one application is CPU bound and does not benefit from larger memory bandwith and the other is memory bound and does not benefit from having more CPU capacity. So different applications affect each other's performance with different coefficients. 
\end{comment}
%%%%%%%%%%%%%%%%%%%%%%%%%%%%%%%%%%%%%%%%%%%%%%%%%%%%%%%%%%%%%%%%%%%%%%%%%%%
%%%%%%%%%%%%%%%%%%%%%%%%%%%%%%%%%%%%%%%%%%%%%%%%%%%%%%%%%%%%%%%%%%%%%%%%%%%
%%%%%%%%%%%%%%%%%%%%%%%%%%%%%%%%%%%%%%%%%%%%%%%%%%%%%%%%%%%%%%%%%%%%%%%%%%%
%%%%%%%%%%%%%%%%%%%%%%%%%%%%%%%%%%%%%%%%%%%%%%%%%%%%%%%%%%%%%%%%%%%%%%%%%%%
%%%%%%%%%%%%%%%%%%%%%%%%%%%%%%%%%%%%%%%%%%%%%%%%%%%%%%%%%%%%%%%%%%%%%%%%%%%
%%%%%%%%%%%%%%%%%%%%%%%%%%%%%%%%%%%%%%%%%%%%%%%%%%%%%%%%%%%%%%%%%%%%%%%%%%%
\indent \textbf{Example:} As an example, suppose we have 5 different applications and 5 different cache levels with different capacities of 128KB, 256KB, 512KB, 1MB and 2MB. In addition, suppose the 128kB cache level can not accomodate more than one application and 256kB cache can accomodate 2 applications, 512kB level can have 4 applications, 1MB cache can have 8 applications and 2MB cache can have at most 16 applications. Let's assume the following matrix be the utility function of each application on each cache level. \\
\indent Some applications may get better utility from smaller cache space since they are less congested and since these applications have low data locality, moving to larger cache spaces not only does not increase their performance but also degrades the performance by evicting other applications from the cache and making contention on the memory bandwidth which is a more vital resource for them \footnote{\textit{libquantum}, \textit{streamL}, \textit{sphinx3}, \textit{lbm} and \textit{mcf} are examples of such applications.}.  \\
\begin{equation}
M = \bordermatrix{~ & 1way & 2way & 4way & 8way & 16way \cr
  App1 & 1.9 & 1.7 & 1.5 & 1 & 0.9 \cr
  App2 & 1.6 & 1.3 & 1.1 & 0.8 & 0.7 \cr
  App3 & 1.4 & 1.0 & 0.6 & 0.5 & 0.4 \cr
  App4 & 0.3 & 0.6 & 0.9 & 1.2 & 1.4 \cr
  App5 & 0.7 & 0.8 & 1.1 & 1.4 & 1.7 \cr}
\end{equation}
%%%%%%%%%%%%%%%%%%%%%%%%%%%%%%%%%%%%%%%%%%%%%%%%%%%%%%%%%%%%%%%%%%%%%%%%%%%
%%%%%%%%%%%%%%%%%%%%%%%%%%%%%%%%%%%%%%%%%%%%%%%%%%%%%%%%%%%%%%%%%%%%%%%%%%%
In the first iteration of the bidding, the first 3 applications bid for the most profitable resource which is 128kB cache and they submit a bid equal to the difference of profit between the first and the second most profitable resource. Therefore, the first application, submits 0.2 bid to 128kb and the second application submits 0.3 and the third application submits 0.4. Since only one of the players can acquire the 128kB cache space, the first application will get it. The 4th and 5th application compete for 2MB cache space and they both get it with the sum bid of both which is 0.5. In the next round, the prices will be updated and since applications 2 and 3 don't have any cache assignment compete for the 256kB cache space and each bid 0.2 which is the difference between 1.7 and 1.5 and 1.3 and 1.1 in the performance matrix accordingly. Since the second level cache can accommodate both applications the price will be updated and the minimum bidding price for someone to get this cache level is updated to the minimum bid of both which is 0.2. Therefore, if some application bid more than 0.2 it can acquire the resource and the application with the smallest bid has to resubmit the bid to acquire the resource. Figure~\ref{fig:first_round}, ~\ref{fig:second_round}, and ~\ref{fig:third_round} show the bidding steps and the prices and minimum price of bidding accordingly. As seen from the figures, the auction terminates in three iterations when there exist five applications. 
%%%%%%%%%%%%%%%%%%%%%%%%%%%%%%%%%%%%%%%%%%%%%%%%%%%%%%%%%%%%%%%%%%%%%%%%%%%
%%%%%%%%%%%%%%%%%%%%%%%%%%%%%%%%%%%%%%%%%%%%%%%%%%%%%%%%%%%%%%%%%%%%%%%%%%%
%%%%%%%%%%%%%%%%%%%%%%%%%%%%%%%%%%%%%%%%%%%%%%%%%%%%%%%%%%%%%%%%%%%%%%%%%%% 
%%%%%%%%%%%%%%%%%%%%%%%%%%%%%%%%%%%%%%%%%%%%%%%%%%%%%%%%%%%%%%%%%%%%%%%%%%%
%%%%%%%%%%%%%%%%%%%%%%%%%%%%%%%%%%%%%%%%%%%%%%%%%%%%%%%%%%%%%%%%%%%%%%%%%%%
\vspace{-1\baselineskip}
\subsection{A Case Study for Hybrid Cache Game}
\begin{figure}[!tb]
\centering
\includegraphics[height=1.5in, width=3.5in]{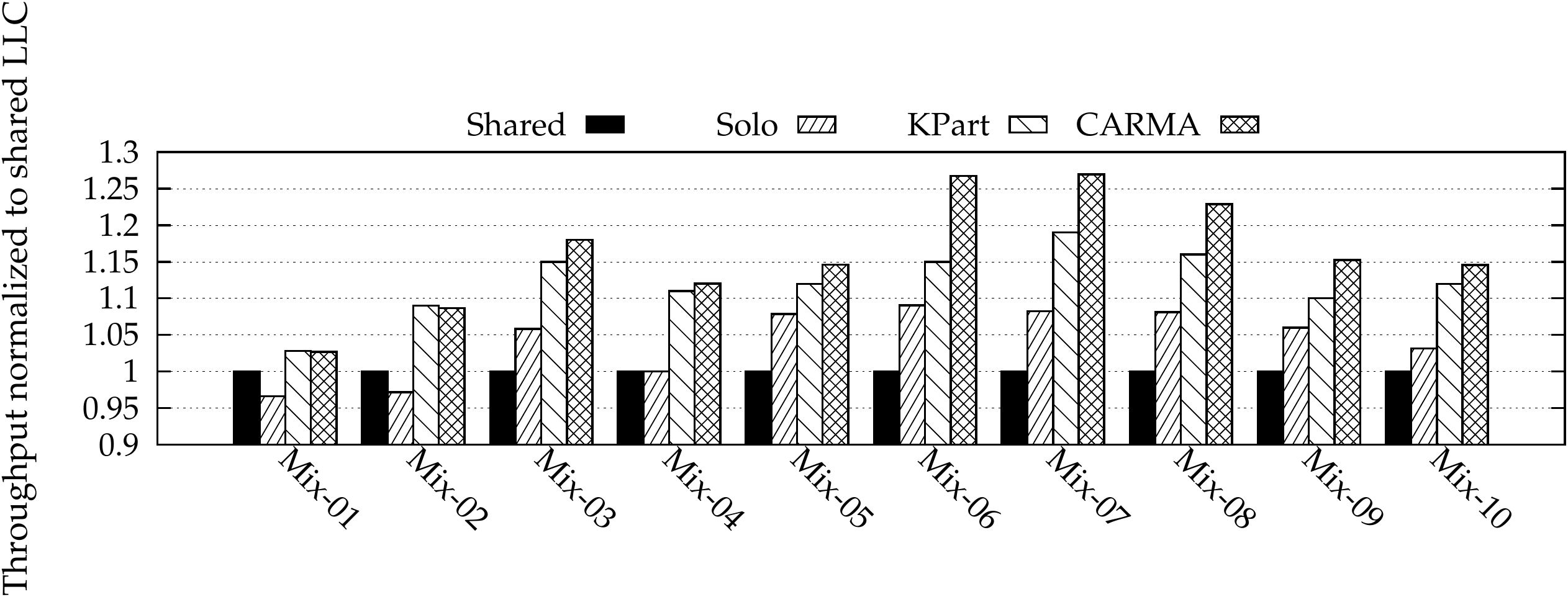}
\caption{Throughput of a shared, solo, CARMA and KPart cache allocation schemes.}
\label{fig:IPC_mix}
\end{figure}

\begin{figure}[!tb]
\centering
\includegraphics[height=1.5in, width=3.5in]{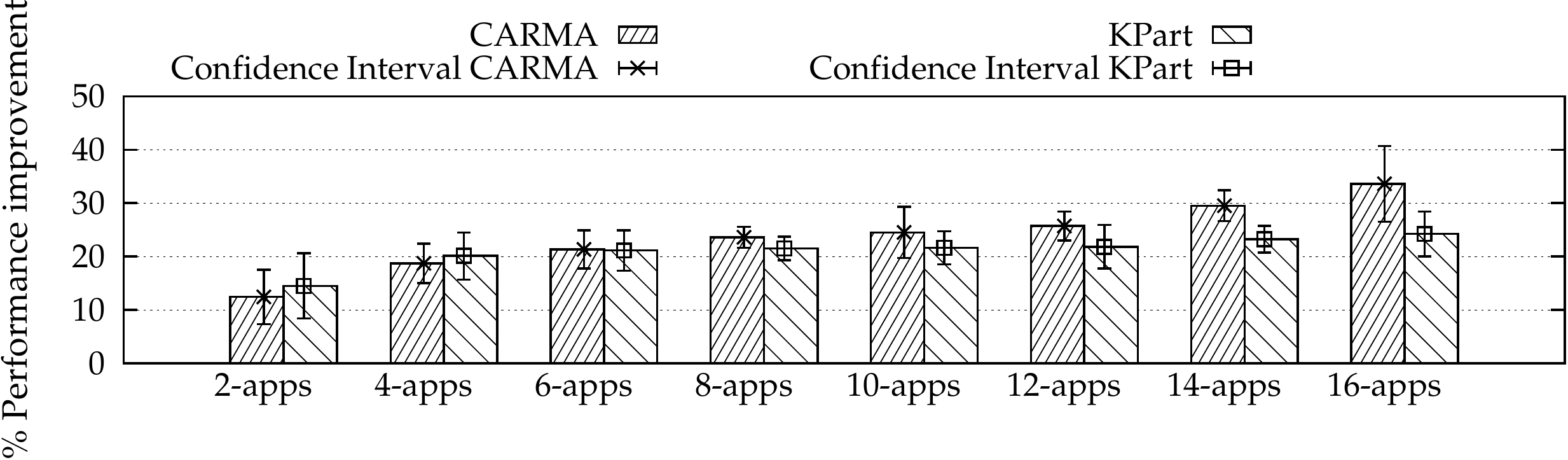}
\caption{Performance improvement of CARMA and KPart for different number of applications with respect to shared LLC for the case study of cache congestion game.}
\label{fig:scalability}
\end{figure}

\begin{figure}[!tb]
\centering
\includegraphics[height=1.5in, width=2.5in]{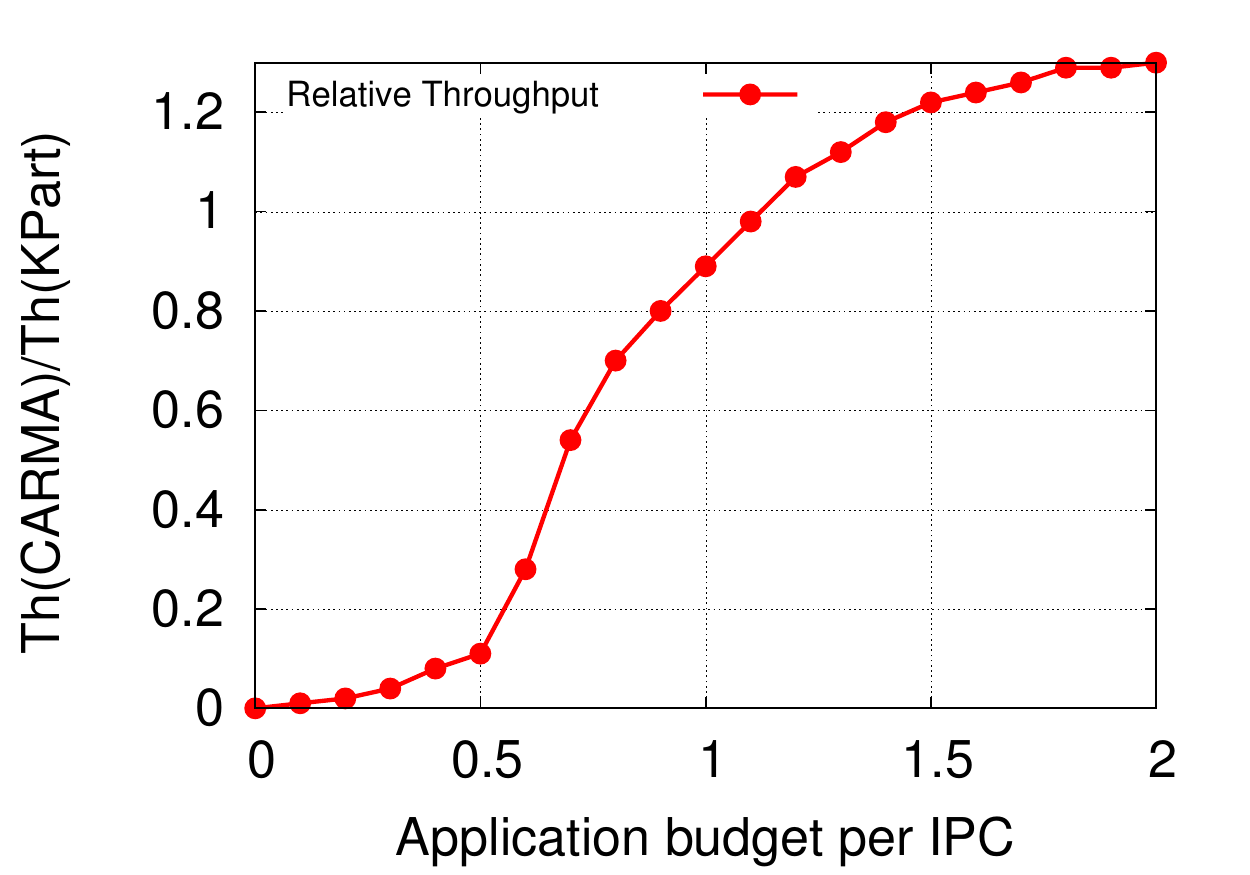}
\caption{Relative throughput w.r.t applications' normalized budget.}
\label{fig:Relative_Thr}
\vspace{-1\baselineskip}
\end{figure}

\indent In hybrid cache game, each cache partition can have a cluster of applications. We use different mixes of 4 to 16 applications from \textit{Spec 2006} to evaluate the performance of our proposed approach compared to others. To evaluate our approach, we selected the state-of-the-art centralized cache partitioner~\cite{8327002} (KPart) as a competitor which aims at maximizing the global IPC speedup. CARMA uses multi-resource valuations, so each application can have any criteria to maximize its payoff. In order to provide a fair comparison with our approach, we use IPC speedup as the optimization goal for all applications.\\% having the profile of the applications and similar bundled resources for the participants.\\ 
\indent Figure~\ref{fig:IPC_mix} shows the normalized throughput of 10 different mix of applications \cite{srikantaiah2011morphcache}, using CARMA, KPart~\cite{8327002}, equal separate cache partitioning and completely shared cache space after convergence. Furthermore, Figure~\ref{fig:scalability} shows the scalability of our proposed algorithm. When the number of co-runners increases from 2 to 16, the performance improves without any need to track each applications' performance in a central module. Having full information about applications' profiles, CARMA outperforms the other centralized competitors, when the number of the applications increases.\\
%The results of throughput and performance compared to DABMFT are pretty the same except little changes because of partitioning is a little sensitive to initial state, but the optimal valuation and allocation after KPart partitioning are the same. It is why that we haven't shown it in the figures not to confuse the other results. However, DABMFT calculation is more complex for computer architecture purpose, since it is designed for multiple applications with multiple sellers.
\indent Since KPart is a centralized (not an auction-based) approach, we assume that it has an unlimited budget. The budget matters in CARMA. We setup another experiment to track the variations of the normalized throughput versus the normalized budget for a mix of 16 applications. Figure~\ref{fig:Relative_Thr} shows that the throughput of CARMA, is very sensitive to the budget. The throughput changes dramatically at some inflection point, and at the end it is saturated but higher than KPart.
%%%%%%%%%%%%%%%%%%%%%%%%%%%%%%%%%%%%%%%%%%%%%%%%%%%%%%%%%%%%%%%%%%%%%%%%%%%
%%%%%%%%%%%%%%%%%%%%%%%%%%%%%%%%%%%%%%%%%%%%%%%%%%%%%%%%%%%%%%%%%%%%%%%%%%%
%%%%%%%%%%%%%%%%%%%%%%%%%%%%%%%%%%%%%%%%%%%%%%%%%%%%%%%%%%%%%%%%%%%%%%%%%%%
%%%%%%%%%%%%%%%%%%%%%%%%%%%%%%%%%%%%%%%%%%%%%%%%%%%%%%%%%%%%%%%%%%%%%%%%%%%
%%%%%%%%%%%%%%%%%%%%%%%%%%%%%%%%%%%%%%%%%%%%%%%%%%%%%%%%%%%%%%%%%%%%%%%%%%%
\section{Related Work} \label{Related_works}
With rapid improvement in computer technology, more and more cores are embedded in a single chip and applications competing for a shared resource is becoming common. On the one hand, managing scheduling of shared resources for a large number of applications is challenging in a sense that the operating system doesn't know what is the performance metric for each application. But on the other hand, the operating system has a global view of the whole state of the system and can guide applications on choosing the shared resources.\\ 
\indent There have been several works, for managing the shared cache in multi-core systems. Qureshi et al. \cite{qureshi2006utility} showed that assigning more cache space to applications with more cache utility does not always lead to better performance since there exist applications with very low cache reuse which may have very high cache utilization. \\
\indent Several software and hardware approaches have been proposed to find the optimal partitioning of cache space for different applications \cite{zhuravlev2010addressing}. However, most of these approaches use brute force search of all possible combinations to find the best cache partitioning in runtime or introduce a lot of overhead. There have been some approaches which use binary search to reduce searching all possible combinations \cite{kim2004fair, lin2008gaining, tam2009rapidmrc}. But none of these methods are scalable for the future many-core processor designs.\\
\indent There exists prior game-theoretic approaches designing a centralized scheduling framework that aims at a fair optimization of applications' utility \cite{zahedi2014ref, llull2017cooper, ghodsi2011dominant, zahedi2015sharing, fan2016computational}. Zahedi et al. in REF \cite{zahedi2014ref, zahedi2015sharing} use the Cobb-Douglas production function as a fair allocator for cache and memory bandwidth. They show that the Cobb-Douglas function provides game-theoretic properties such as sharing incentives, envy-freedom, and Pareto efficiency. But their approach is still centralized and spatially divides the shared resources to enforce a fair near-optimal policy sacrificing the performance. In their approach, the centralized scheduler assumes all applications have the same priority for cache and memory bandwidth, while we do not have any assumption on this. Further, our auction-based resource allocation can be used for any number of resources and any priority for each application and the centralized scheduler does not need to have a global knowledge of these priorities.  \\
\indent Ghodsi et al. in DRF \cite{ghodsi2011dominant} use another centralized fair policy to maximize the dominant resource utilization. But in practice, it is not possible to clone any number of instances of each resource. %  the underlying scenario cloning the instances is very limited or superficial for practical purposes. 
Cooper \cite{llull2017cooper} enhances REF to capture colocated applications fairly, but it only addresses the special case of having two sets of applications with matched resources. Fan et al. \cite{fan2016computational} exploits computational sprinting architecture to improve task throughput assuming a class of applications where boosting their performance by increasing the power. \\
\indent While all prior works use a centralized scheduling that provides fairness and assumes the same utility function for all, co-runners might have completely diverse needs and it is not efficient to use the same fairness/performance policy across them. 
Our auction-based resource scheduling provides scalability since individual applications compete for the shared resources based on their utility and the burden of decision making is removed from the central scheduler. We believe that future CMPs should move toward a more decentralized approach which is more scalable and provides a fair allocation of resources based on the applications' needs. \\ 
%Providing the scalability of the system is getting better if we make the individual applications our self-administrators, and we can remove most of the performance-restricting policies such as fairness constraints from the centralized decision-maker. \\
\indent Auction theory which is a subfield of economics has recently been used as a tool to solve large-scale resource assignment in cloud computing \cite{krishna2009auction, parsons2011auctions}. In an auction process, the buyers submit bids to get the commodities and sellers want to sell their commodities with the maximum price as possible. Also auction-based allocators ~\cite{Zhang2017DAB,8004498} are multi-buyers with multi-seller but there is only one resource to bid. So, they cannot be used for our purpose, since we have only one seller with multiple bundled resources. That is why we choose a simpler related scheme for a computer architecture to get higher performance with lower transactions and auctions. \\
\indent Our auction-based algorithm is inspired by work of Bertsekas \cite{bertsekas1998network} that uses an auction-based approach for network flow problems. Our algorithm is an extension of local assignment problem proposed by Bertsekas et al. that has been shown to converge to the global assignment within a linear approximation.
%\textcolor{red}{Not Complete yet}
%\vspace{-1\baselineskip}
\vspace{-1\baselineskip}
\section{Conclusion} 
\label{Conclusion}
%\vspace{-1\baselineskip}
%\begin{normalsize}
This paper proposes a distributed resource allocation approach for large-scale servers. The traditional resource management system is not scalable, especially when tracking the application's dynamic behavior. The main cause of this complexity is the centralized decision making which leads to higher time and space complexity. With increasing number of cores per chip, the scalability of assigning different resources to different applications becomes more challenging in future generation CMP systems. In addition, diversity in application's need makes a single objective function inefficient to get an optimal and fair performance metric. We introduce a framework to map the allocation problem to the known auction economy model where the applications compete for the shared resources based on their utility metrics of interest.
%\end{normalsize}
%\textcolor{red}{Not Complete yet}\\
%\vspace{-1\baselineskip}

%\vspace{-0.5\baselineskip}
%\input{Sections/Introduction}
%\vspace{-0.5\baselineskip}
%\input{Sections/Motivation}
%\vspace{-0.5\baselineskip}
%\input{Sections/Case_Studies}
%\vspace{-0.5\baselineskip}
%\input{Sections/Related_Works}
%\vspace{-0.5\baselineskip}
%\input{Sections/Conclusion}
%\vspace{-0.5\baselineskip}
%\input{Sections/Appendix}
%\vspace{-0.5\baselineskip}

%%%%%%% -- PAPER CONTENT ENDS -- %%%%%%%%

%%%%%%%%% -- BIB STYLE AND FILE -- %%%%%%%%
%\bibliographystyle{ieeetr}
%\bibliography{ref}
%%%%%%%%%%%%%%%%%%%%%%%%%%%%%%%%%%%%
\bibliographystyle{unsrt}{%\small  %unsrt
\bibliography{bib/IEEEconf}

\begin{thebibliography}{10}

\bibitem{tootaghajICCD}
D.~Z. Tootaghaj and F.~Farhat.
\newblock Cage: A contention-aware game-theoretic model for heterogenous
  resource assignment.
\newblock In {\em The 35th IEEE International Conference on Computer Design
  (ICCD)}, 2017.

\bibitem{tang2011impact}
L.~Tang, J.~Mars, N.~Vachharajani, R.~Hundt, and M.~L. Soffa.
\newblock The impact of memory subsystem resource sharing on datacenter
  applications.
\newblock In {\em Computer Architecture (ISCA), 2011 38th Annual International
  Symposium on}, 2011.

\bibitem{zhuravlev2010addressing}
S.~Zhuravlev, S.~Blagodurov, and A.~Fedorova.
\newblock Addressing shared resource contention in multicore processors via
  scheduling.
\newblock In {\em ACM SIGARCH Computer Architecture News}. ACM, 2010.

\bibitem{hsu2006communist}
L.~R. Hsu, S.~K. Reinhardt, R.~Iyer, and S.~Makineni.
\newblock Communist, utilitarian, and capitalist cache policies on cmps: caches
  as a shared resource.
\newblock In {\em PACT}. ACM, 2006.

\bibitem{kim2004fair}
S.~Kim, D.~Chandra, and Y.~Solihin.
\newblock Fair cache sharing and partitioning in a chip multiprocessor
  architecture.
\newblock In {\em Proceedings of the 13th International Conference on Parallel
  Architectures and Compilation Techniques}. IEEE Computer Society, 2004.

\bibitem{cho2006managing}
S.~Cho and L.~Jin.
\newblock Managing distributed, shared l2 caches through os-level page
  allocation.
\newblock In {\em MICRO}, 2006.

\bibitem{farhat2016stochastic}
F.~Farhat, D.~Z. Tootaghaj, Y.~He, A.~Sivasubramaniam, M.~T. Kandemir, and
  C.~R. Das.
\newblock Stochastic modeling and optimization of stragglers.
\newblock {\em IEEE Transactions on Cloud Computing}, 2016.

\bibitem{tootaghaj2016optimal}
D.~Z. Tootaghaj and F.~Farhat.
\newblock Optimal placement of cores, caches and memory controllers in network
  on-chip.
\newblock {\em arXiv preprint arXiv:1607.04298}, 2016.

\bibitem{tootaghaj2015evaluating}
D.~Z. Tootaghaj, F.~Farhat, M.~Arjomand, P.~Faraboschi, M.t~T. Kandemir,
  A.~Sivasubramaniam, and C.~R. Das.
\newblock Evaluating the combined impact of node architecture and cloud
  workload characteristics on network traffic and performance/cost.
\newblock In {\em IEEE international symposium on Workload characterization
  (IISWC)}. IEEE, 2015.

\bibitem{farhat2016towardsStoc}
F.~Farhat, D.~Z. Tootaghaj, and M.~Arjomand.
\newblock Towards stochastically optimizing data computing flows.
\newblock {\em arXiv preprint arXiv:1607.04334}, 2016.

\bibitem{tootaghaj2015thesis}
D.~Z. Tootaghaj.
\newblock {\em Evaluating Cloud Workload Characteristics}.
\newblock PhD thesis, Pennsylvania State University, 2015.

\bibitem{liu2004organizing}
C.~Liu, A.~Sivasubramaniam, and M.~T. Kandemir.
\newblock Organizing the last line of defense before hitting the memory wall
  for cmps.
\newblock In {\em IEEE Proceedings on Software}, 2004.

\bibitem{qureshi2006utility}
Y.~N.~Patt M.~K.~Qureshi.
\newblock Utility-based cache partitioning: A low-overhead, high-performance,
  runtime mechanism to partition shared caches.
\newblock In {\em MICRO}. IEEE Computer Society, 2006.

\bibitem{lin2008gaining}
J.~Lin, Q.~Lu, X.~Ding, Z.~Zhang, X.~Zhang, and P.~Sadayappan.
\newblock Gaining insights into multicore cache partitioning: Bridging the gap
  between simulation and real systems.
\newblock In {\em IEEE 14th International Symposium on High Performance
  Computer Architecture (HPCA)}, pages 367--378, 2008.

\bibitem{iyer2004cqos}
R.~Iyer.
\newblock Cqos: a framework for enabling qos in shared caches of cmp platforms.
\newblock In {\em ICS}. ACM, 2004.

\bibitem{rafique2006architectural}
N.~Rafique, W.~T. Lim, and M.~Thottethodi.
\newblock Architectural support for operating system-driven cmp cache
  management.
\newblock In {\em PACT}. ACM, 2006.

\bibitem{jiang2008analysis}
Y.~Jiang, X.~Shen, J.~Chen, and R.~Tripathi.
\newblock Analysis and approximation of optimal co-scheduling on chip
  multiprocessors.
\newblock In {\em PACT}. ACM, 2008.

\bibitem{yekkehkhany2017gb}
A.~Yekkehkhany, A.~Hojjati, and M.~H. Hajiesmaili.
\newblock Gb-pandas: Throughput and heavy-traffic optimality analysis for
  affinity scheduling.
\newblock In {\em IFIP Proceedings of Performance (IFIP Performance 2017)},
  2017.

\bibitem{xie2016scheduling}
Q.~Xie, A.~Yekkehkhany, and Y.~Lu.
\newblock Scheduling with multi-level data locality: Throughput and
  heavy-traffic optimality.
\newblock In {\em INFOCOM}. IEEE, 2016.

\bibitem{zhou2014sharing}
Y.~Zhou and D.~Wentzlaff.
\newblock The sharing architecture: sub-core configurability for iaas clouds.
\newblock In {\em ACM SIGARCH Computer Architecture News}. ACM, 2014.

\bibitem{tootaghaj2011game}
D.~Z. Tootaghaj, F.~Farhat, M.~R. Pakravan, and M.~R. Aref.
\newblock Game-theoretic approach to mitigate packet dropping in wireless
  ad-hoc networks.
\newblock In {\em IEEE CCNC}, 2011.

\bibitem{tootaghaj2011risk}
D.~Z. Tootaghaj, F.~Farhat, M.~R. Pakravan, and M.~R. Aref.
\newblock Risk of attack coefficient effect on availability of ad-hoc networks.
\newblock In {\em IEEE CCNC}, 2011.

\bibitem{kotobi2017spectrum}
K.~Kotobi and S.~G. Bilen.
\newblock Spectrum sharing via hybrid cognitive players evaluated by an m/d/1
  queuing model.
\newblock {\em EURASIP Journal on Wireless Communications and Networking},
  2017.

\bibitem{kotobi2015introduction}
K.~Kotobi and S.~G. Bilen.
\newblock Introduction of vigilante players in cognitive networks with moving
  greedy players.
\newblock In {\em Vehicular Technology Conference (VTC Fall)}. IEEE, 2015.

\bibitem{kesidis2013distributed}
G.~Kesidis, K.~Kotobi, and C.~Griffin.
\newblock Distributed aloha game with partially rule-based cooperative, greedy,
  and vigilante players.
\newblock {\em Department of Computer Science and Engineering, Penn State
  University, Tech. Rep. CSE}, 2013.

\bibitem{kurve2013agent}
A.~Kurve, K.~Kotobi, and G.~Kesidis.
\newblock An agent-based framework for performance modeling of an optimistic
  parallel discrete event simulator.
\newblock {\em Complex Adaptive Systems Modeling}, 2013.

\bibitem{wang2017using}
N.~Nasiriani, C.~Wang, G.~Kesidis, and B.~Urgaonkar.
\newblock Using burstable instances in the public cloud: Why, when and how?
\newblock {\em Proceedings of the ACM on Measurement and Analysis of Computing
  Systems}, 2017.

\bibitem{wang2015recouping}
C.~Wang, N.~Nasiriani, G.~Kesidis, B.~Urgaonkar, Q.~Wang, L.~Y. Chen, A.~Gupta,
  and R.~Birke.
\newblock Recouping energy costs from cloud tenants: Tenant demand response
  aware pricing design.
\newblock In {\em Proceedings of the 2015 ACM Sixth International Conference on
  Future Energy Systems}. ACM, 2015.

\bibitem{osborne1994course}
M.~J. Osborne and A.~Rubinstein.
\newblock {\em A course in game theory}.
\newblock MIT press, 1994.

\bibitem{zahedi2014ref}
S.~M. Zahedi and B.~C. Lee.
\newblock Ref: Resource elasticity fairness with sharing incentives for
  multiprocessors.
\newblock {\em ACM SIGARCH Computer Architecture News}, 2014.

\bibitem{llull2017cooper}
Q.~Llull, S.~Fan, S.~M. Zahedi, and B.~C. Lee.
\newblock Cooper: Task colocation with cooperative games.
\newblock In {\em IEEE International Symposium on High Performance Computer
  Architecture (HPCA)}. IEEE, 2017.

\bibitem{ghodsi2011dominant}
A.~Ghodsi, M.~Zaharia, B.~Hindman, A.~Konwinski, S.~Shenker, and I.~Stoica.
\newblock Dominant resource fairness: Fair allocation of multiple resource
  types.
\newblock In {\em NSDI}, 2011.

\bibitem{zahedi2015sharing}
S.~M. Zahedi and B.~C. Lee.
\newblock Sharing incentives and fair division for multiprocessors.
\newblock {\em IEEE Micro}, 2015.

\bibitem{fan2016computational}
S.~Fan, S.~M. Zahedi, and B.~C. Lee.
\newblock The computational sprinting game.
\newblock In {\em ACM SIGOPS Operating Systems Review}. ACM, 2016.

\bibitem{bertsekas1998network}
D.~P. Bertsekas.
\newblock {\em Network Optimization: continuous and discrete methods}.
\newblock Athena Scientific, 1998.

\bibitem{kyle1985continuous}
A.~S. Kyle.
\newblock Continuous auctions and insider trading.
\newblock {\em Econometrica: Journal of the Econometric Society}, 1985.

\bibitem{vasconcelos2009bipartite}
Cristina~Nader Vasconcelos and Bodo Rosenhahn.
\newblock Bipartite graph matching computation on gpu.
\newblock In {\em EMMCVPR}. Springer, 2009.

\bibitem{ferber1999multi}
J.~Ferber.
\newblock {\em Multi-agent systems: an introduction to distributed artificial
  intelligence}, volume~1.
\newblock Addison-Wesley Reading, 1999.

\bibitem{mantevo}
\texttt{http:/manetovo.org}.

\bibitem{Spec:website}
\texttt{http://www.spec.org/spec2006}.

\bibitem{binkert2011gem5}
N.~Binkert, B.~Beckmann, G.~Black, S.~K. Reinhardt, A.~Saidi, A.~Basu,
  J.~Hestness, D.~R. Hower, T.~Krishna, S.~Sardashti, et~al.
\newblock The gem5 simulator.
\newblock {\em ACM SIGARCH Computer Architecture News}, 2011.

\bibitem{Gem5:website}
\texttt{http://gem5.org/}.

\bibitem{suh2002new}
G.~E. Suh, S.~Devadas, and L.~Rudolph.
\newblock A new memory monitoring scheme for memory-aware scheduling and
  partitioning.
\newblock In {\em High-Performance Computer Architecture, 2002. Proceedings.
  Eighth International Symposium on}. IEEE, 2002.

\bibitem{suh2004dynamic}
G.~E. Suh, L.~Rudolph, and S.~Devadas.
\newblock Dynamic partitioning of shared cache memory.
\newblock {\em The Journal of Supercomputing}, 2004.

\bibitem{tam2009rapidmrc}
D.~K Tam, R.~Azimi, L.~B. Soares, and M.~Stumm.
\newblock Rapidmrc: approximating l2 miss rate curves on commodity systems for
  online optimizations.
\newblock In {\em ACM SIGARCH Computer Architecture News}, 2009.

\bibitem{krishna2009auction}
Krishna. V.
\newblock {\em Auction theory}.
\newblock Academic press, 2009.

\bibitem{parsons2011auctions}
S.~Parsons, J.~A. Rodriguez-Aguilar, and M.~Klein.
\newblock Auctions and bidding: A guide for computer scientists.
\newblock {\em ACM Computing Surveys (CSUR)}, 2011.

\end{thebibliography}
} 
%%%%%%%%%%%%%%%%%%%%%%%%%%%%%%%%%%%%%%%%%%%%%%%%%%%%%%%%%%%%%%%%%%%%%%%%%%%%%%%%%%%%%%%%%%
\begin{IEEEbiography}[{\includegraphics[width=1in,height=1.25in,clip,keepaspectratio]{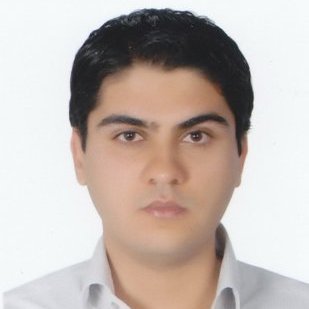}}]{Farshid Farhat} is a PhD candidate at the School of Electrical Engineering and Computer Science, The Pennsylvania State University. He obtained his B.Sc., M.Sc., and Ph.D. degrees in Electrical Engineering from Sharif University of Technology, Tehran, Iran. His current research interests include resource allocation in parallel, distributed systems, computer vision and image processing. He is working on the modeling and analysis of image composition and aesthetics
using deep learning and image retrieval on different platforms ranging from smartphones to high-performance
computing clusters.%received his B.S. and M.S. degrees in electrical engineering from Sharif University of Technology. He is currently a Ph.D candidate at the EECS department, Pennsylvania State University. His current research interests include resource allocation in parallel and distributed systems.  
\end{IEEEbiography}
%%%%%%%%%%%%%%%%%%%%%%%%%%%%%%%%%%%%%%%%%%%%%%%%%%%%%%%%%%%%%%%%%%%%%%%%%%%%%%%%%%%%%%%%%%%
%\input{Sections/Appendix}
%%%%%%%%%%%%%%%%%%%%%%%%%%%%%%%%%%%%%%%%%%%%%%%%%%%%%%%%%%%%%%%%%%%%%%%%%%%%%%%%%%%%%%%%%%
\begin{IEEEbiography}[{\includegraphics[width=1in,height=1.25in,clip,keepaspectratio]{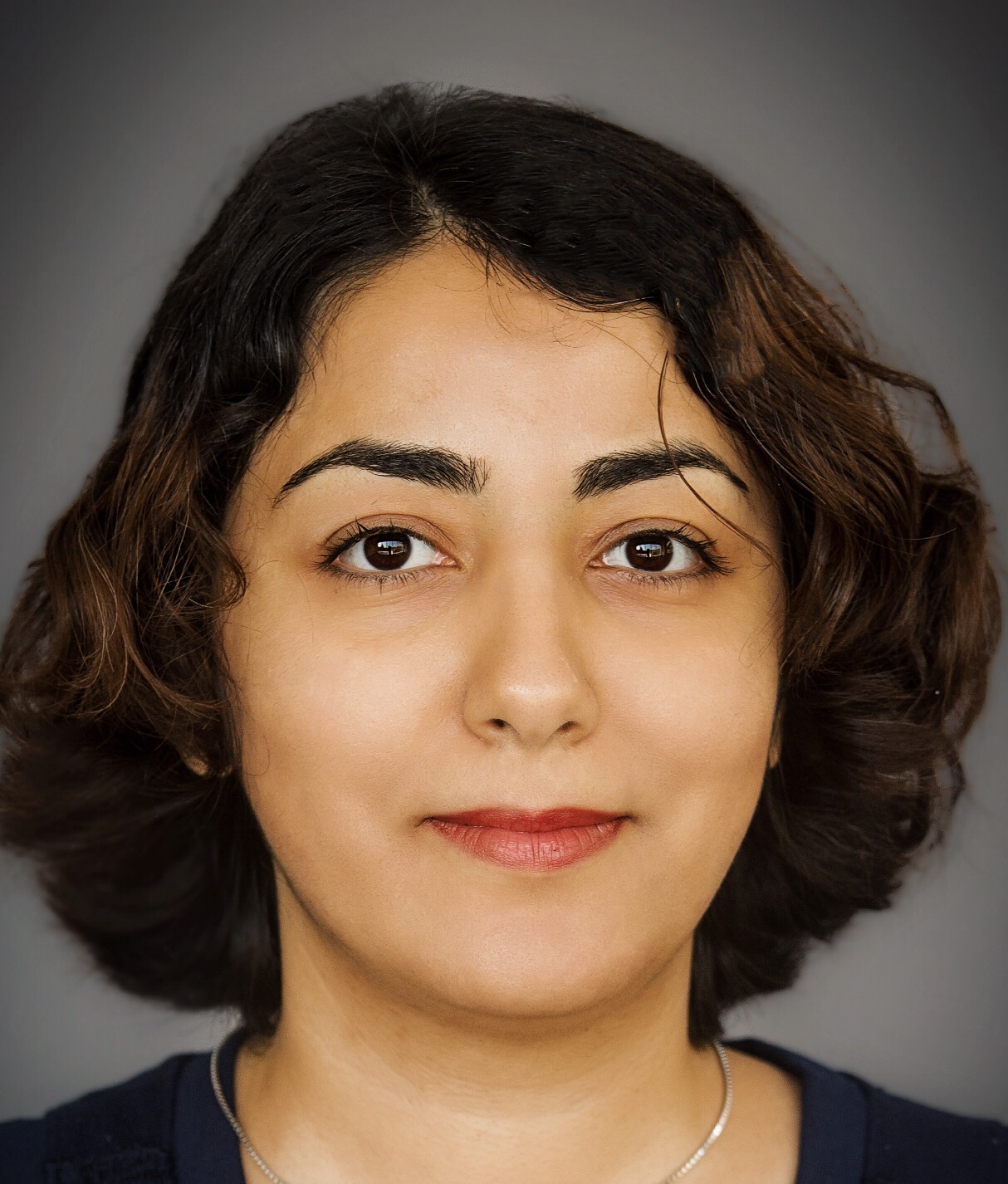}}]{Diman Zad Tootaghaj} is a Ph.D. student in the department of computer science and engineering at the Pennsylvania State University. She received B.S. and M.S. degrees in Electrical Engineering from Sharif University of Technology, Iran in 2008 and 2011 and a M.S. degree in Computer Science and Engineering from the Pennsylvania State University in 2015. She is a member of Institute for Networking and Security Research (INSR) under supervision of Prof. Thomas La Porta (advisor), Dr. Ting He (co-advisor), and Dr. Novella Bartolini. Her current research interests include computer network, recovery approaches, distributed systems, and stochastic analysis.
\end{IEEEbiography}

\end{document}